\pdfoutput=1

\documentclass[11pt,twoside,a4paper,cmspaper,final,collab]{cms-tdr}
\def\svnVersion{307613}\def\cmsCernNo{2013-037}\def\cmsCernDate{\today}\def\cmsMessage{Published in the Journal of High Energy Physics as \href{http://dx.doi.org/10.1007/JHEP09(2015)201}{\doi{10.1007/JHEP09(2015)201.}}}

\begin{document}\cmsNoteHeader{EXO-12-007}

\hyphenation{had-ron-i-za-tion}
\hyphenation{cal-or-i-me-ter}
\hyphenation{de-vices}
\RCS$HeadURL: svn+ssh://svn.cern.ch/reps/tdr2/papers/EXO-12-007/trunk/EXO-12-007.tex $
\RCS$Id: EXO-12-007.tex 299469 2015-08-09 11:36:15Z alverson $
\cmsNoteHeader{EXO-12-007}

\title{Search for neutral color-octet weak-triplet scalar particles in proton-proton collisions at $\sqrt{s} = 8$\TeV}

\date{\today}

\abstract{
A search for pair production of neutral color-octet weak-triplet scalar particles ($\Theta^{0}$) is performed in processes where one $\Theta^{0}$ decays to a pair of b quark jets and the other to a Z boson plus a jet, with the Z boson decaying to a pair of electrons or muons.
The search is performed with data collected by the CMS experiment at the CERN LHC corresponding to an integrated luminosity of 19.7\fbinv of proton-proton collisions at $\sqrt{s} = 8$\TeV.
The number of observed events is found to be in agreement with the standard model predictions.
The 95\% confidence level upper limit on the product of the cross section and branching fraction is obtained as a function of the $\Theta^{0}$ mass.
The 95\% confidence level lower bounds on the $\Theta^{0}$ mass are found to be 623 and 426\GeV, for two different octo-triplet theoretical scenarios.
These are the first direct experimental bounds on particles predicted by the octo-triplet model.
}

\hypersetup{%
pdfauthor={CMS Collaboration},%
pdftitle={Search for neutral color-octet weak-triplet scalar particles in proton-proton collisions at sqrt(s) = 8 TeV},
pdfsubject={CMS},%
pdfkeywords={CMS, physics, exotica, Z+jets}}

\maketitle

\section{Introduction}\label{sec:intro}
The CERN LHC is well suited to searches for new colored particles.
These searches have targeted particles including vector-like quarks (color triplet spin-1/2 particles), colorons (color octet spin-1 particles), gluinos (color octet Majorana fermions), and color-octet weak-singlet scalars~\cite{tagkey2014149, atlasDijet, Chatrchyan:2013izb, Aad201222, Chatrchyan:2014lfa, atlasGluino}.
With the recent discovery of a Higgs boson~\cite{Aad20121,Chatrchyan201230, 2014arXiv1412.8662C},
so far consistent with being a neutral scalar colorless
particle~\cite{tagkey2013120,PhysRevD.89.092007},
it is also interesting to look for additional scalars, in this case with color charge, with masses close to the electroweak scale.
Particles with this combination of spin, charge, and color quantum numbers have not been thoroughly sought at collider experiments.
Neutral color-octet spin-0 particles ($\Theta^0$), for example, emerge in the octo-triplet model~\cite{2012PhRvD..85g5020D},
which includes three particles
($\Theta^{+}, \Theta^{0}, \Theta^{-}$) that transform as a color octet and weak triplet under $SU(3)_{c} \times SU(2)_{W}$.
The $\Theta^{0}$ may arise from the sector responsible for breaking extended gauge symmetry associated with colorons~\cite{Hall1985430,Hill1991419, Bai:2010dj}.
Octo-triplets may also be fermion-antifermion bound states~\cite{Bai:2010mn, Bai:2010qg}, or elementary particles from non-minimal grand unified theory.
To date, no direct experimental bound has been set for particles predicted by the octo-triplet model.

Octo-triplet particles would be produced in pairs at the LHC either through
quark-antiquark annihilation or gluon-gluon interactions, with the former mediated by coloron (G') particles or by gluons (\Glu).
If a G' is produced, it can decay to $\Theta^{0}\Theta^{0}$, $\Theta^{0}\phi_\mathrm{I}$, or two quarks,
where $\phi_\mathrm{I}$ is a color singlet from a general renormalizable coloron model~\cite{Bai:2010dj}.
In the present analysis we are searching for $\Theta^{0}$ produced in pairs.
The $\phi_\mathrm{I}$ mass ($m_{\phi_\mathrm{I}}$) is a free parameter and has an indirect impact on the $\Theta^{0}$ pair production
cross section, via the G' width.

Unlike color-octet weak-singlet particles, octo-triplet particles do not decay to gluons.
However, with a $\Theta^{0}$ in the loop, one-loop decays to an electroweak boson and a gluon are allowed.
There is a potentially similar rate of decays to standard model (SM) quark pairs via mixing through a vector-like quark, which can be as heavy as a few\TeV.
These heavy vector-like quarks appear in many extensions of the SM, such
as composite Higgs models~\cite{PhysRevD.59.075003} and little Higgs models~\cite{PhysRevD.67.095004}.
In the former, the Higgs boson may be a bound state of a top quark and the heavy vector-like quark,
where the binding is provided by the spin-1 G' coloron.

\begin{figure}[hb]
  \centering
    \includegraphics[width=0.48\textwidth]{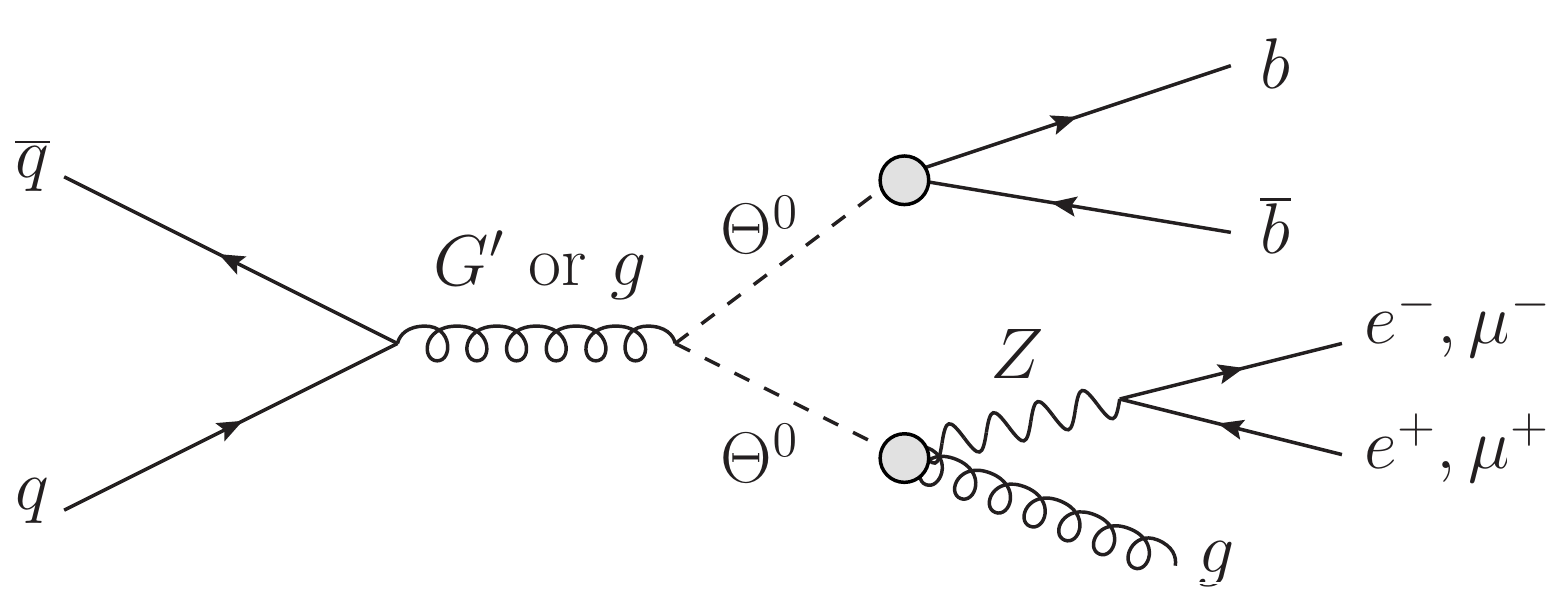}
    \includegraphics[width=0.48\textwidth]{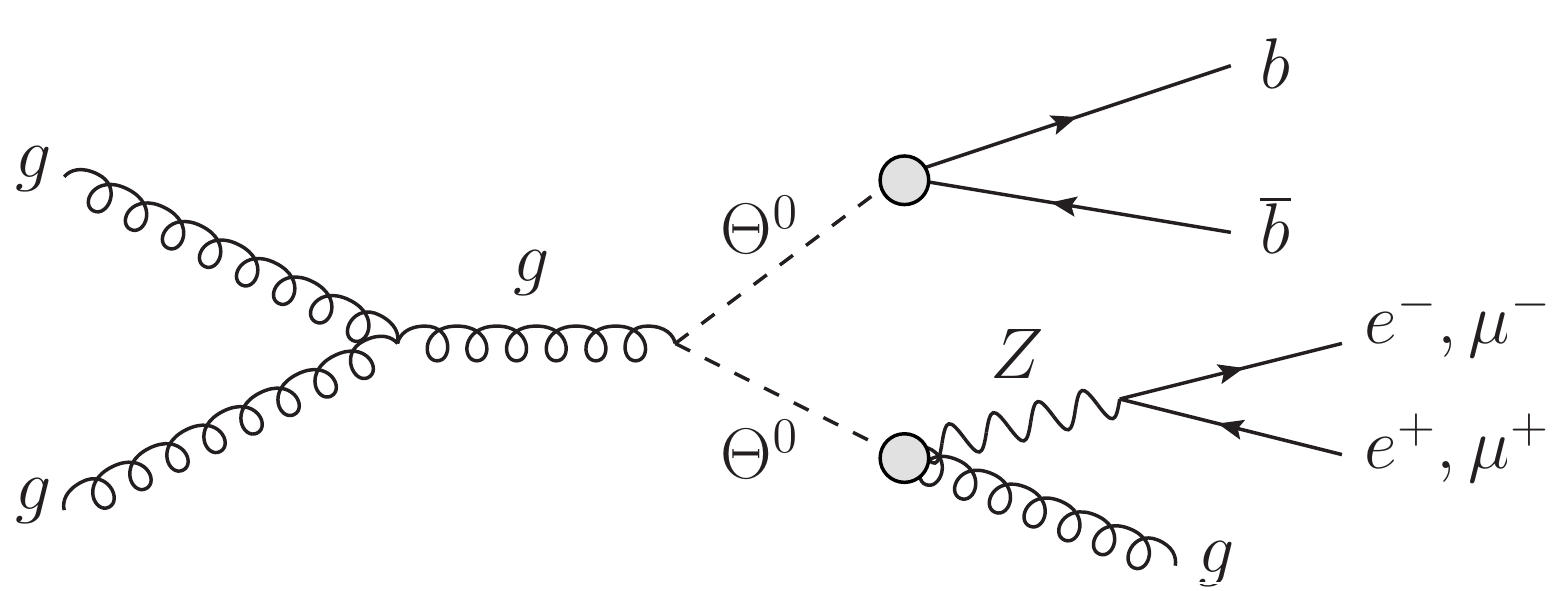}
  \caption[Theta0 pair production and decay]{Leading order Feynman
diagrams for $\Theta^{0}$ pair production by quark-antiquark annihilation, through
an intermediate G' or gluon (left), and the gluon-gluon interaction (right). In addition to the $s$-channel gluon process shown in the figure, the gluon-gluon interaction can also proceed directly through $gg\to\Theta^0\Theta^0$ or a $t$-channel $\Theta^0$ exchange.
The decays of the $\Theta^{0}$ and \Z boson are also included, where the $\Theta^{0}$ decays are described in the main text.
}
  \label{fig:production}
\end{figure}

In this paper we assume that one $\Theta^{0}$ decays to two b quark jets and the other to a \Z boson plus a jet originating from a gluon, with $\Z \to \ell\ell$, where $\ell$ is an electron or muon.
With no existing constraints on the ratio of $\Theta^{0}$  decays to quarks versus those decaying to $\Z\Glu$, we have studied a final state that includes both decay modes.
Requiring $\Z \to \ell\ell$ provides a strong experimental signature for event selection and background suppression.
The leading order (LO) Feynman diagrams for $\Theta^{0}$ pair production by quark-antiquark annihilation
and gluon-gluon interaction, and the decays explored in this analysis, are shown in Fig.~\ref{fig:production}.

\begin{figure}[htb]
  \centering
  \includegraphics[width=0.6\textwidth]{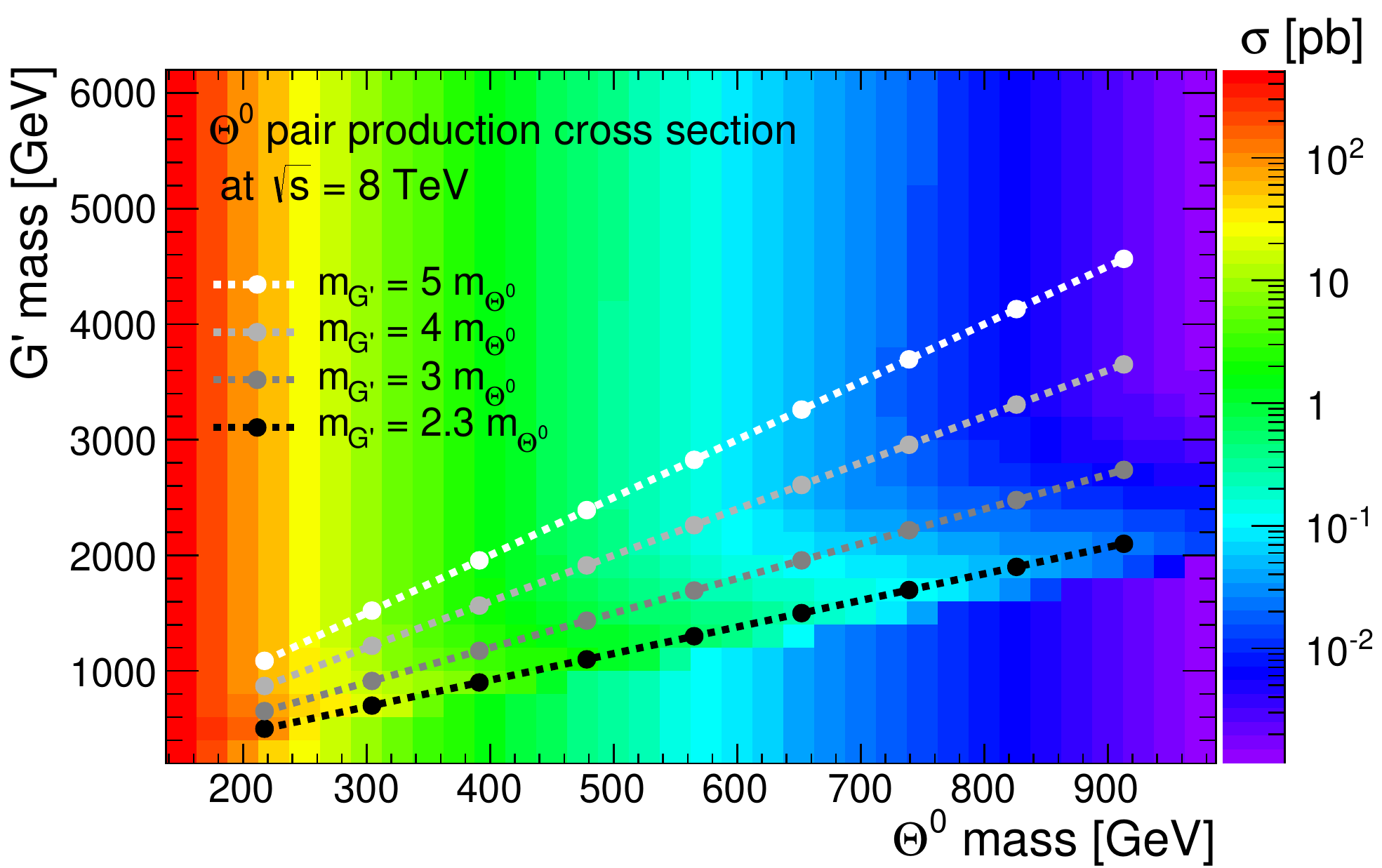}
  \caption[Cross section scan]{Cross sections for $\Theta^0$ pair production, as a function of the coloron $\mathrm{G'}$ and the octo-triplet $\Theta^0$ masses. The dotted lines show a few examples of assumed linear relationship between the two masses.}
  \label{fig:twoD_scan_wide_zoom}
\end{figure}

The signal production cross section for the octo-triplet model~\cite{2012PhRvD..85g5020D} is evaluated at LO with the \MADGRAPH v4.5.1 program~\cite{MadGraph}.
Figure~\ref{fig:twoD_scan_wide_zoom} shows the values of the cross section in the 2-dimensional plane
of G' mass ($m_{\mathrm{G'}}$) and $\Theta^{0}$ mass ($m_{\Theta^{0}}$).
The masses of $\Theta^{0}$, G', and $\phi_{\rm{I}}$ are free parameters in general, but we use $m_{\phi_{\rm{I}}} = 125\GeV + m_{\Theta^{0}} / 3$ as a benchmark.
We explore two distinct mass scenarios: $m_{\mathrm{G'}} = 2.3 m_{\Theta^{0}}$ and $m_{\mathrm{G'}} = 5 m_{\Theta^{0}}$.
The first example corresponds to a scenario in which the signal cross section is enhanced because of the additional contributions from colorons (see Fig.~\ref{fig:twoD_scan_wide_zoom}), while the second relation corresponds to a scenario in which the contribution from colorons is negligible.
For both mass relations we use $\Theta^{0}$ masses, ranging from 200 to 900\GeV, which are large enough to allow $\Theta^{0}$ decay
to on-shell $\Z$ bosons, but small enough to be within the sensitivity of this search.
The first mass relation, $m_{\mathrm{G'}} = 2.3 m_{\Theta^{0}}$, results in $\Theta^{0}$ pair production cross sections from
125\unit{pb} down to 2.23$\times 10^{-2}$\unit{pb}, for the range of $\Theta^{0}$ masses considered.
This mass relation also sets the G' mass sufficiently above $\Theta^{0}$ pair production threshold,
but small enough compared to $m_{\Theta^{0}}$ that production through G'
is considerable, at 49 to 86\% of the total, depending on $m_{\Theta^{0}}$.
In this case, quark-antiquark annihilation dominates,
while gluon-gluon interaction production decreases with increasing $\Theta^{0}$ (and G') mass,
because of the differences in the parton distribution functions (PDF) of quarks
and gluons \cite{Campbell:2006wx}.
The second mass relation considered, $m_{\mathrm{G'}} = 5 m_{\Theta^{0}}$,
results in $\Theta^{0}$ pair production cross sections from 63.3\unit{pb} down to 2.62$\times 10^{-3}$\unit{pb}, for
the range of $\Theta^{0}$ masses considered, which are a factor of two smaller
than for the first mass relation.
This corresponds to a region where gluon-gluon interactions dominate and production through G' is just 0.6 to 3.6\% of the total, depending on the $\Theta^{0}$ mass.
The cross sections discussed above agree with those calculated in Ref.~\cite{Chivukula:2013hga} where
a similar model is considered.

The decay modes $\Theta^{0} \to \Z\Glu$ and $\Theta^{0} \to \gamma\Glu$ are invariant under $SU(2)_{W}$.
However, in the Lagrangian, the coefficients of these modes turn out to be much smaller than one.
This feature motivates consideration of other decay modes, which can be represented by higher-dimensional operators, and therefore are usually considered as negligible with respect to dominant decay modes.
In particular, $q\bar{q}$ decay modes could be as large as the electroweak boson plus gluon modes \cite{2012PhRvD..85g5020D}.
In the $\Theta^{0}$ mass range we consider, the dominant fermion decay mode is either \bbbar or \ttbar depending on the values of coefficients appearing in different terms of the  $\Theta^{0} \to \PQq\PAQq$ Lagrangian.
As a benchmark, we assign half of the branching fraction to $\Theta^{0} \to \bbbar$, leaving the other half to the $\Theta^{0} \to \Z\Glu$ and $\Theta^{0} \to \gamma\Glu$ decay modes (\ie $\mathcal{B}(\Theta^{0} \to \ttbar) = 0$).
The ratio of branching fractions $\mathcal{B}(\Theta^{0} \to \Z\Glu)/\mathcal{B}(\Theta^{0} \to \gamma\Glu)$ is set by the mass of the $\Theta^{0}$.
For the range of $\Theta^{0}$ masses explored, $\mathcal{B}(\Theta^{0} \to \Z\Glu)$ is  to $38\%$ and $\mathcal{B}(\Theta^{0} \to \gamma \Glu)$ accounts for the remaining  to $17\%$.
The total cross section times branching fraction for the final state studied
in this analysis ranges from 2.77\unit{pb} down to $5.74\times10^{-4}$\unit{pb} for the mass relation
$m_{\mathrm{G'}} = 2.3 m_{\Theta^{0}}$, and 1.40\unit{pb} down to $6.73\times10^{-5}$\unit{pb} for the mass relation
$m_{\mathrm{G'}} = 5 m_{\Theta^{0}}$.
\section{CMS detector}

The central feature of the CMS apparatus is a superconducting solenoid of 6\unit{m} internal
diameter, providing a magnetic field of 3.8\unit{T}. Within the solenoid volume are a silicon pixel
and strip tracker, a lead tungstate crystal electromagnetic calorimeter, and a brass and scintillator hadron
calorimeter, each composed of a barrel and two endcap sections. Muons are measured in gas-ionization detectors embedded in the steel flux-return yoke outside the solenoid. Extensive
forward calorimetry complements the coverage provided by the barrel and endcap detectors.
A more detailed description of the CMS detector, together with a definition of the coordinate system used and the relevant kinematic variables, can be found in Ref.~\cite{Chatrchyan:2008zzk}.

\section{Event selection}
\label{sec:selection}

This analysis uses proton-proton collision data at a center-of-mass energy of 8\TeV collected
with dilepton ($\Pe\Pe$, $\mu\mu$, and $\Pe\mu$) triggers
with transverse momentum $\pt$ thresholds of 17 and 8\GeV for the two lepton candidates.
The $\Pe\mu$ data set is used to estimate \ttbar background.
The data sample corresponds to an integrated luminosity of 19.7\fbinv.
We select events with a final state containing a Z boson and a jet ($\Z+\text{jet}$),
together with two b quark jets (b jet pair), where the Z boson decays to a pair of electrons or muons.

Electrons are reconstructed using selection criteria that take into account radiated photons~\cite{Khachatryan:2015hwa}.
Candidates are required to have $\PT > 20\GeV$ and $\abs{\eta} < 2.5$ where $\eta$ is the pseudorapidity.
Electrons from photon conversions are rejected.
An isolation condition is imposed within a cone of $\Delta R = \sqrt{\smash[b]{(\Delta\eta)^2 + (\Delta\phi)^2} }< 0.3$
around the electron.
The sum of the \pt of charged particles and transverse energy $\et =E\sin\theta$ of neutral particles within this cone is corrected in each event for energy deposits due to additional interactions within beam bunch crossings (pileup).
The corrected sum must be less than 15\% of the electron \pt.

Muons are reconstructed using selection criteria based on quantities measured in the
tracker and muon sub-detectors.  Candidates are required to have $\PT > 20\GeV$
and $\abs{\eta} < 2.4$.
The track associated with the muon candidate is required
to have hits in the pixel, silicon strip, and muon systems.
An isolation condition requires that the sum of the \pt of charged particles and \et of neutral particles
within a cone of $\Delta R < 0.4$ around the muon is less than 12\% of the muon \pt.
The pileup correction for muon isolation is similar to the one applied for electrons.

A particle-flow technique \cite{pflow, CMS:2010byl} is used to identify jet constituents, which are input to the \FASTJET algorithm \cite{Cacciari:2005hq,Cacciari:2011ma} for clustering using the anti-\kt algorithm \cite{2008JHEP...04..063C} with a distance parameter of 0.5.
The jet energy scale (JES) is measured in data with $\Z/\gamma+$jet and dijet events \cite{Chatrchyan:1369486, CMS-DP-2013-033}, and a correction is applied to both data and simulated samples.
The corrected jets must have $\PT > 40\GeV$ and $\abs{\eta} < 2.4$.
Jets within $\Delta R = 0.3$ from an isolated electron or muon as defined above are not counted as part of the 3 jet requirement.
Additional requirements, based on the energy balance between charged and neutral
hadronic energy in the jet, are used to reduce contamination from misidentified
jets.  A jet is considered b tagged, consistent with originating from the
hadronization of a bottom quark, if it satisfies the ``loose operating point''
requirements of the combined secondary vertex tagger~\cite{2013JInst...8P4013T}.
The tagging efficiency is about 70--90\% and mistagging rate is about 10--20\%, depending on \pt and $\eta$ of the jets.

Candidate events must satisfy the following criteria: at least one reconstructed primary vertex satisfying $\abs{z}<24$\unit{cm} and impact parameter less than 2\unit{cm};
an opposite-sign, same-flavor lepton (electron or muon) pair, sharing a primary
vertex and with an invariant mass
between  80 and 100\GeV to form a Z boson candidate;
the two leptons separated from jets by $\Delta R > 0.5$;
at least two b-tagged jets forming a b jet pair system (sum of the b jet four-momenta); and at least one additional jet.
If more than two b-tagged jets are present in the event, the jets with the largest \pt values
are selected for the b jet pair system.

The mass of one $\Theta^{0}$ candidate is reconstructed from the b jet pair system, and the mass of the other
$\Theta^{0}$ candidate is reconstructed from the combination of the Z boson candidate and the highest
$\PT$ jet that is not a part of the b jet pair system.
With this prescription, the correct combination of the \Z boson and the jet in simulated signal samples is chosen
about 65 to 80\% of the time, depending on $m_{\Theta^{0}}$.

\begin{figure}[htb]
  \centering
  \includegraphics[width=0.48\textwidth]{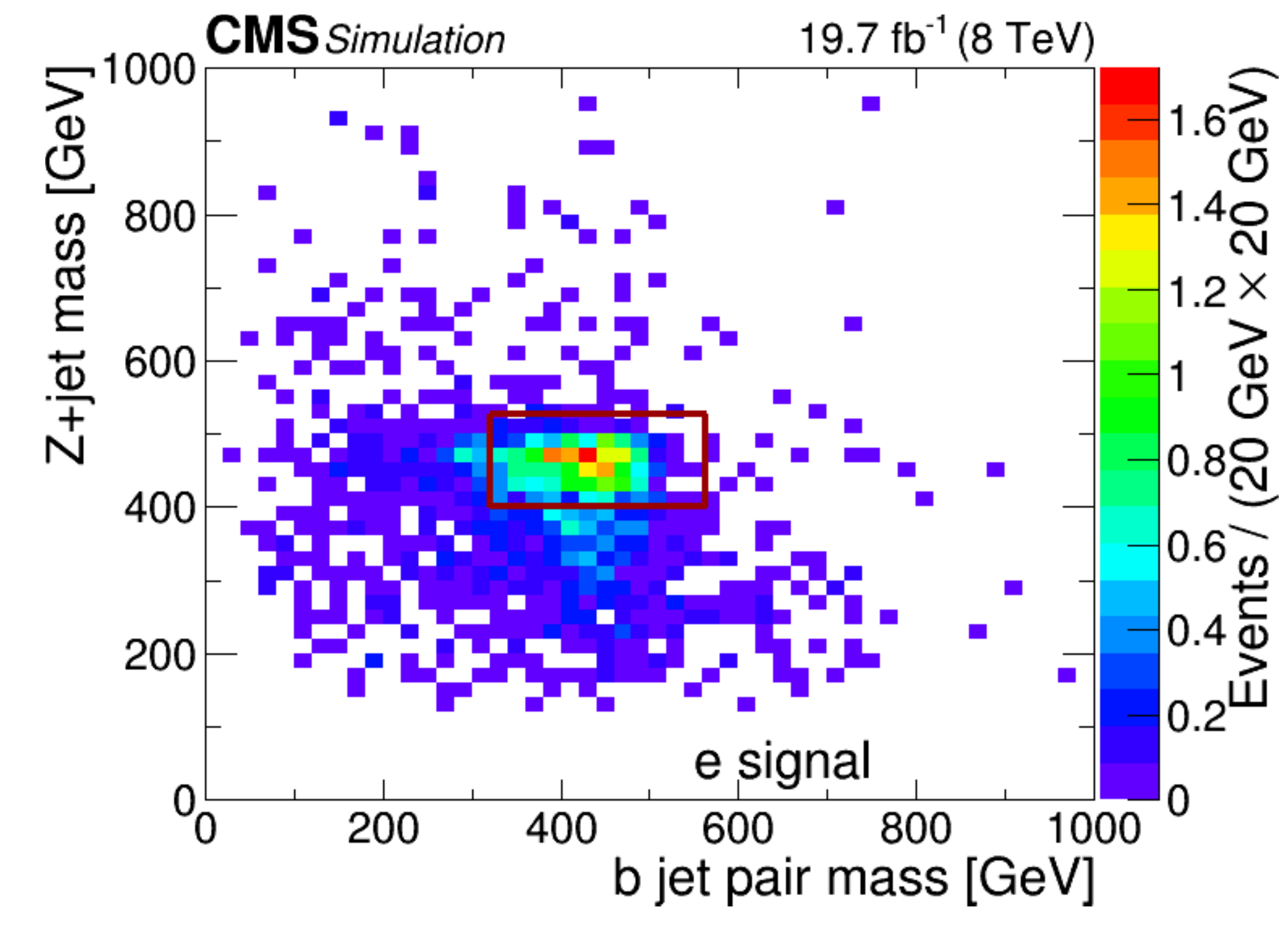}
  \includegraphics[width=0.48\textwidth]{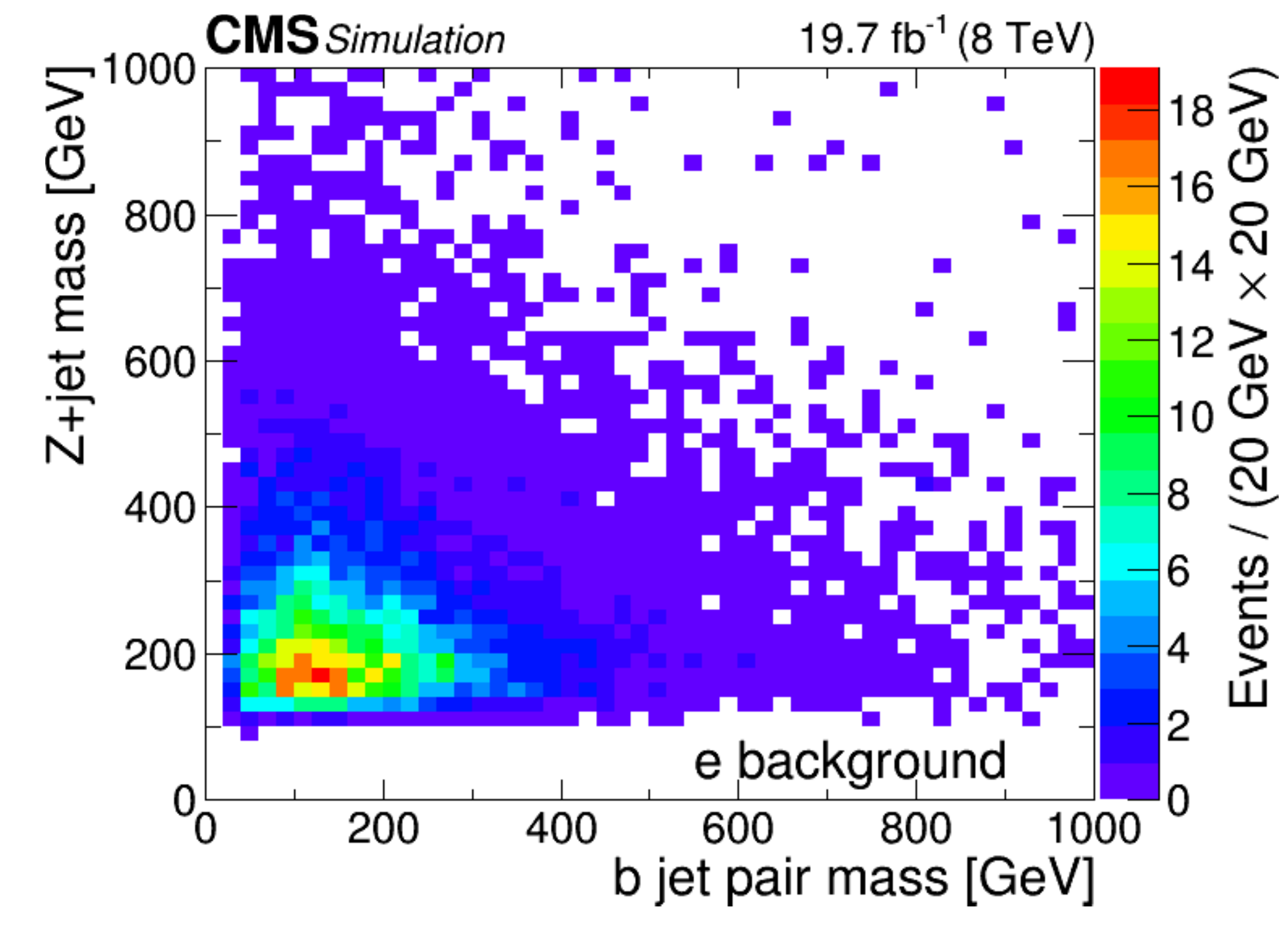}
  \caption{Distributions of $\Z+\text{jet}$ mass versus b jet pair mass for signal events with $m_{G'} = 1100\GeV$ and $m_{\Theta^{0}} = 478\GeV$ (left) and background events (right), in the electron channel (distributions are similar in the muon channel).
The open red rectangle on left indicates the signal region, as described in the text.}
  \label{fig:twoDmass}
\end{figure}

For each hypothetical mass point, we consider signal events
in a rectangle that is formed in the two-dimensional mass plane defined by the two reconstructed $\Theta^0$ masses.
We determine the center, length, and width of the rectangle for each mass point by fitting the b jet pair and $\Z+\text{jet}$
mass distributions in the simulated signal samples with a modified Gaussian function that has an additional parameter to account for low-mass non-Gaussian tails.
The length and width of each rectangle are defined as $\pm 3\sigma_{\mathrm{sd}}$ where $\sigma_{\mathrm{sd}}$ is the standard deviation from the modified Gaussian fit.
For example, the rectangle is 240\GeV long and 130\GeV wide for the signal with $m_{G'} = 1100\GeV$ and $m_{\Theta^{0}} = 478\GeV$ (Fig.~\ref{fig:twoDmass} (left)).
We choose a 3$\sigma_{\mathrm{sd}}$ wide window to keep more signal events in middle--high mass regions, since most of the SM background events appear in low-mass region (Fig.~\ref{fig:twoDmass} (right)).
The signal search regions are defined for the electron and muon channels separately, although these turn out to be very similar.

\section{Backgrounds}\label{sec:backgrounds}

The dominant background processes are $\Z+\text{jets}$ and $\ttbar+\text{jets}$.
The contributions from other processes, including diboson+jets and multijet, are small.
The samples used for all background processes are generated at tree level
with the \MADGRAPH program, interfaced with \PYTHIA v6.4~\cite{2006JHEP...05..026S} for showering and hadronization.
The $\Z+\text{jets}$ background is normalized to the next-to-next-to-leading-order cross section using \FEWZ v2.1~\cite{Gavin20112388}.
The $\ttbar+\text{jets}$ and tW+jets (single top quark) backgrounds are normalized to the next-to-next-to-leading-logarithm cross sections~\cite{Czakon:2013goa, PhysRevD.81.054028}.
The diboson+jets backgrounds are normalized to the next-to-leading-order (NLO) cross section from the \MCFM v5.8~\cite{Campbell:2010ff} calculation.
Full simulation of the CMS detector is implemented
using the \GEANTfour package~\cite{Agostinelli2003250}.

\subsection{\texorpdfstring{$\Z+\text{jets}$}{Z+jets}}
More than half of the background events come from Drell--Yan (DY) production in association with jets.
There are two points in estimating the background that require special attention.
First, care must be taken in estimating the yield of $\Z+{\ge}\text{3 jets}$ events as the simulation may not correctly model the kinematic properties of multijets.
Second, the event yield and shape in the $\Z+\PQb$-jets process, where the kinematic properties of b jet pairs and the fraction of heavy-flavor jets in inclusive $\Z+\text{jets}$ might also suffer from mis-modeling in the simulation.
The following paragraphs describe how the data are used to improve on the estimates from simulations.

\begin{figure}[htb]
\captionsetup[subfloat]{labelformat=empty}
  \centering
  \includegraphics[width=0.48\textwidth]{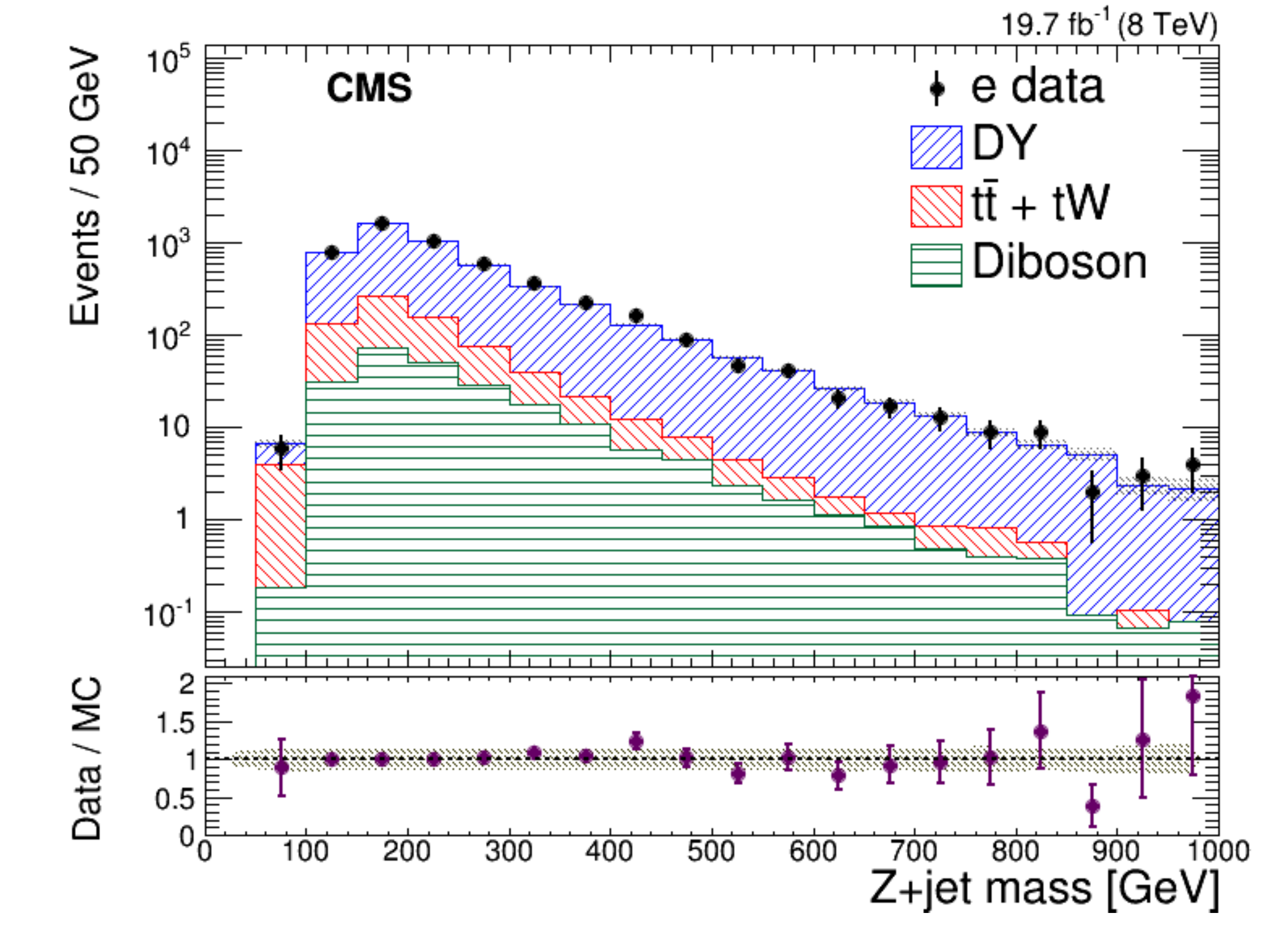}
  \includegraphics[width=0.48\textwidth]{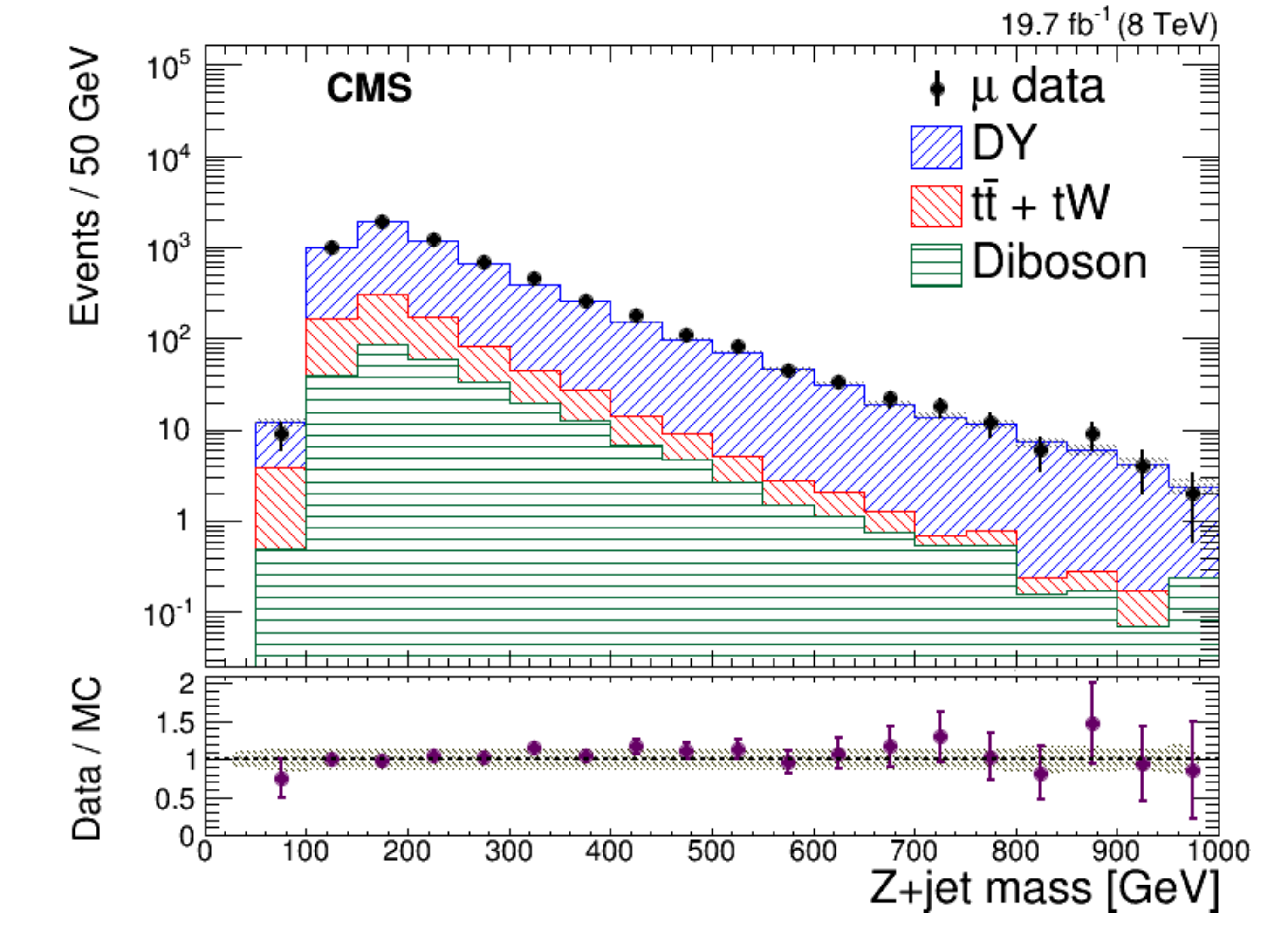}
  \caption{The $\Z+\text{jet}$ mass distribution in the $\Z+{\ge}\text{3 jet}$ (one b tag) control region after the appropriate correction factor has been applied, for the electron channel (left) and muon channel (right).
The panels at the bottom show the ratio of data to background simulation, with the band representing the systematic uncertainty including the normalization uncertainty obtained from the $\Z+{\ge}\text{3 jet}$ (no b tag) control region. }
  \label{fig:dijetCR}
\end{figure}

To address the first point, the event yield from the simulation is multiplied by a correction factor to
normalize it to data in a control region.  This control region is defined such that
the dilepton mass is between 80 and 100\GeV and there are at least three jets, none of which is b tagged.
The correction factor is found to be $0.98 \pm 0.12$ ($0.91 \pm 0.12$) for the electron (muon) channel.
The correction is applied to all signal rectangle regions.
The 12\% uncertainty comes from b tagging, JES, and pileup systematic uncertainties in the control region, added in quadrature,
while the statistical uncertainty is negligible.
The $\Z+\text{jet}$ mass in the $\Z+{\ge}\text{3 jets}$ (one b tag) control region is plotted in Fig.~\ref{fig:dijetCR} after the correction factor has been applied.
Agreement between the data and simulated sample is observed in the $\Z+{\ge}\text{3 jets}$ (no b tag) and $\Z+{\ge}\text{3 jets}$ (one b tag) control regions.

The second point requires a different approach since control regions that include two or more b jets
may suffer from signal contamination.
We take a two-step approach.
First, the simulated $\Z+\PQb$-jets events are weighted by the ratio of the k-factor (NLO cross section divided by LO cross section) for $\Z+2$\,b-jets to that of $\Z+$2 jets, using \MCFM \cite{2003PhRvD..68i4021C}.
The ratios vary from 1.09 to 2.56 in the b jet pair mass range of 20\GeV to 1.8\TeV.
In the simulation, about 20\% of $\Z+\text{jets}$ events in the signal region have at least two jets originating from b quarks.
In the second step, the remaining difference between data and simulation is evaluated as a function of b jet pair mass in the non-signal, off-diagonal regions (sidebands) in the b jet pair mass and $\Z+\text{jet}$ mass plane.
The uncertainty in the $\Z+$heavy-flavor jets processes is taken from this difference or from the uncertainty in the CMS cross section measurement~\cite{Chatrchyan:2014dha}, whichever is larger.
The uncertainty varies from 20 to 50\%.

The uncertainty in the shape of the other variable that is used to define the signal search regions, the $\Z+\text{jet}$ mass distribution, is estimated by comparing the distribution in the simulated $\Z+\text{jets}$ sample with those of several \MADGRAPH samples that are produced with factorization or renormalization scales and matrix element parton shower matching thresholds varied up and down by a factor of two.
The maximum difference between the nominal distribution and the varied distributions is taken as an uncertainty.
The uncertainty varies from 2 to 55\%.

\subsection{\texorpdfstring{$\ttbar+\text{jets}$ and $\PQt\PW+\text{jets}$}{ttbar+jets and tW+jets} }
After accounting for the $\Z+\text{jets}$ background, most of the remaining background events come from $\ttbar+\text{jets}$, with a smaller contribution from tW+jets production.

\begin{figure}[htb]
  \centering
 \includegraphics[width=0.48\textwidth]{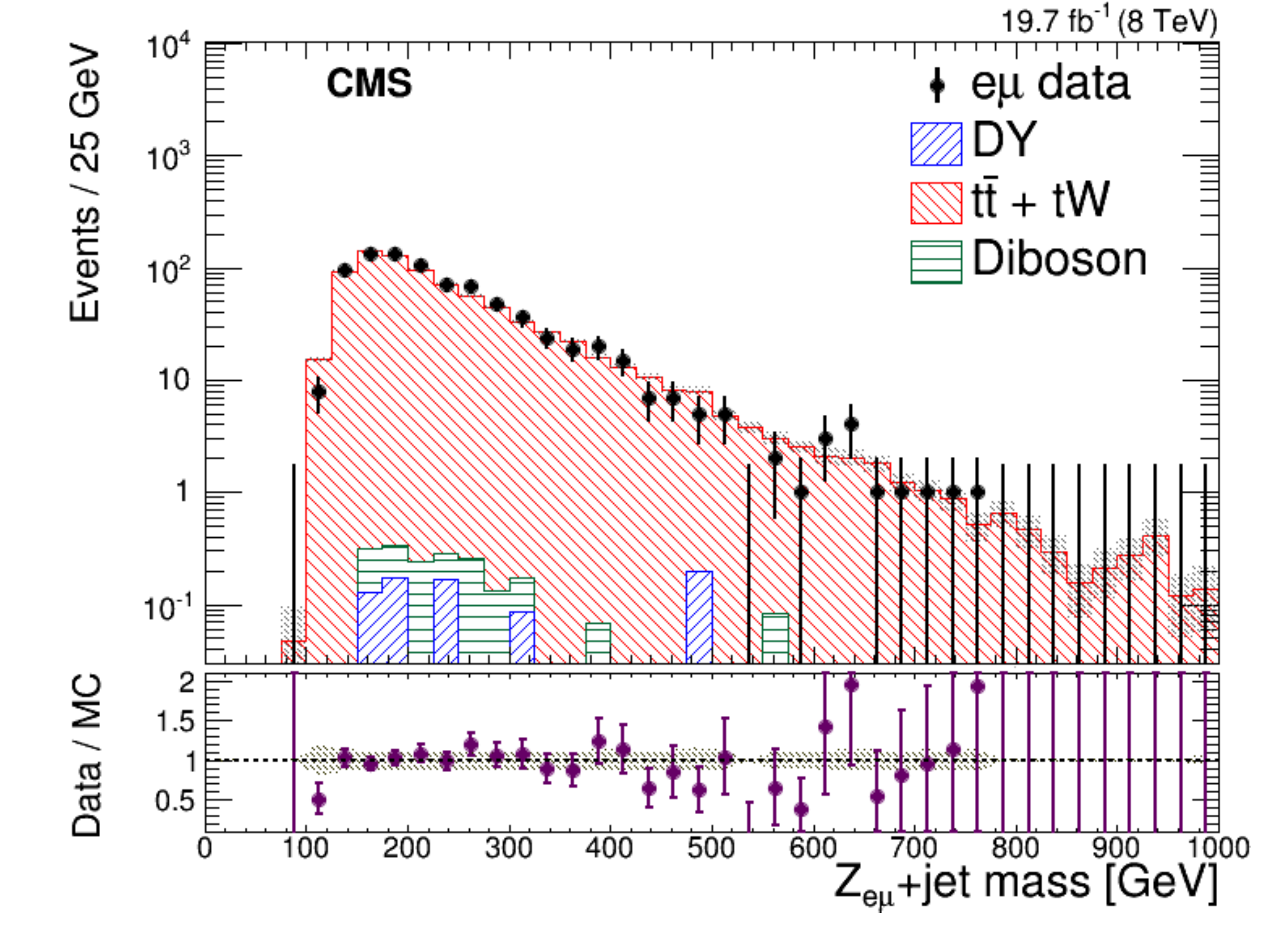}
 \includegraphics[width=0.48\textwidth]{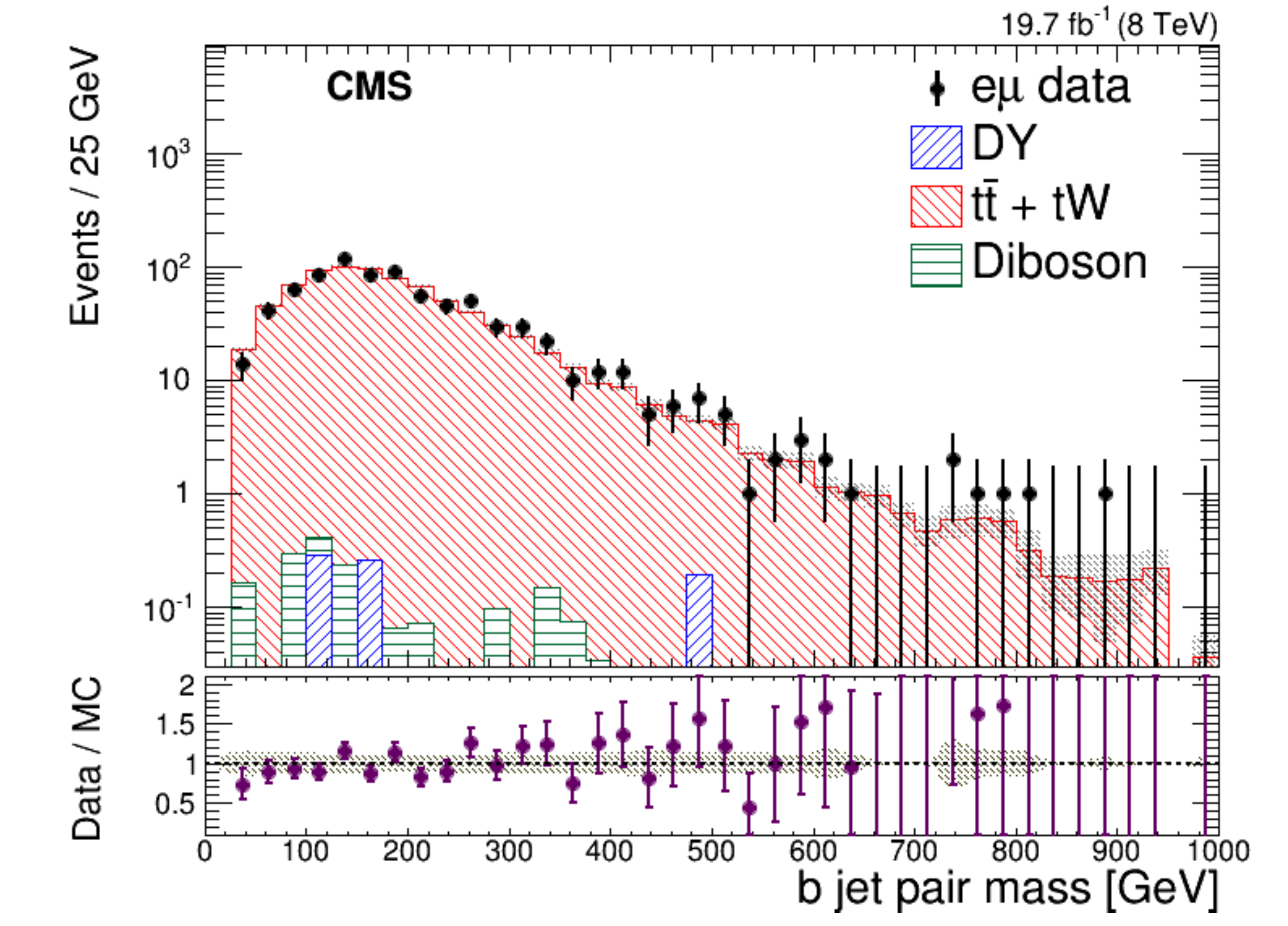}
  \caption{
The $\Z_{\Pe\mu}$+jet (left) and b jet pair mass (right) distributions in the \ttbar control region after the appropriate correction factor is applied.
The panels at the bottom show the ratio of data to background simulation, with the band representing the systematic uncertainty. }
  \label{fig:ttCRmass}
\end{figure}

Both $\ttbar+\text{jets}$ and tW+jets processes can yield final states containing an opposite-sign $\Pe\mu$ pair.
In this analysis, we are considering only opposite-sign, same-flavor lepton pairs, and thus the data
containing an opposite-sign $\Pe\mu$ pair can be used to calibrate the $\ttbar+\text{jets}$ and tW+jets backgrounds.
The overall event yield in the simulation is multiplied by a factor of $1.07 \pm 0.13$ to normalize it to data in the control region containing an $\Pe\mu$ pair with the mass between 60 and 120\GeV ($\Z_{\Pe\mu}$) and at least three jets, at least two of which are b tagged.
The uncertainty comes from b tagging, JES, and pileup systematic uncertainties in the control region, added in quadrature,
while the statistical uncertainty is negligible.
The b jet pair and $\Z_{\Pe\mu}$+jet mass distributions in this control region are plotted in Fig.~\ref{fig:ttCRmass} after the normalization factor is applied.

\subsection{Diboson+jets}
A few percent of the background events come from diboson+jets processes.
The simulated diboson cross sections are found to be in
agreement with those measured by CMS~\cite{2013PhLB..721..190C, Chatrchyan:2014aqa}.
Therefore we make no correction to the event yields.

\subsection{Multijet events}
The estimation of the multijet background with simulated samples is difficult given that the mis-identification rates of jets as electrons or muons are small~\cite{2012JInst...7P0002T, Khachatryan:2015hwa}.
Unlike $\Z \to \ell\ell$ and $\PWm\PWp\to \ell\ell \nu\bar{\nu}$ processes,
there are about equal amount of like-sign and opposite-sign lepton pairs in the background because the leptons are either misidentified or are non-prompt.
In this analysis, we consider only opposite-sign, same-flavor lepton pairs, and thus the data containing like-sign, same-flavor lepton pairs can be used as a control sample to determine the size of the contribution from multijet events.
After the \Z boson mass window selection and the requirement of at least three jets, with at least
two of them being b tagged, and after subtracting events from all other simulated background processes,
27 (13) dielectron (dimuon) events are left in the like-sign
dilepton channel.  This corresponds to 1.6\% (0.7\%) of the number of opposite-sign dilepton events.
These events are distributed mostly in the lower mass region, as is the case for other backgrounds.
Given the small number of like-sign events in the control sample and their mass distribution, the multijet background is found to be negligible.
\section{Systematic uncertainties}
\label{sec:systematics}
We organize the sources of systematic uncertainties in three categories: sources affecting both the signal and background; sources affecting only the background; and sources affecting only the signal.
A summary of the systematic uncertainties is given in Table~\ref{ta:sys290}.

Sources affecting both the signal and background:
\begin{itemize}

\item \textit{b tagging} \\
The data-to-simulation correction factor for b tagging is varied by its uncertainties.
The resulting uncertainties in the number of predicted signal and background events range from 13 to 25\%.

\item \textit {Jet energy scale} \\
The jet energies are varied by the uncertainty in the applied JES correction.
The resulting event yield uncertainties range from 0.2 to 2.6\% for the signal and 4 to 8\% for the background.

\item \textit {Lepton trigger and identification (ID)} \\
The lepton trigger and ID (including isolation) uncertainty is estimated by varying the data-to-simulation correction factor by its uncertainty, which is measured in dilepton data where at least one of the leptons passes stringent identification criteria.
The yield uncertainty is about 1\%, both for the signal and the background.

\item \textit {Pileup modeling}  \\
An uncertainty in the pileup modeling in the signal and background samples is estimated by varying by 5\% the total inelastic cross section, as measured at the LHC~\cite{2011NatCo...2E.463T, tagkey20135}.
The event yield uncertainty is about 1\%, somewhat higher for the signal than for the background.

\end{itemize}

Sources affecting only the background:
\begin{itemize}

\item \textit {$\Z+\text{jet}$ mass shape for $\Z+\text{jets}$ background} \\
The uncertainty is evaluated by comparing the varied simulated samples as mentioned in Section~\ref{sec:backgrounds}.
The shape of the $\Z+\text{jet}$ mass distribution is much less certain in the high-mass tail than in the low-mass region because of the limited number of simulated events with high masses.
The uncertainty in the estimated number of $\Z+\text{jets}$ events varies from 2 to 55\%.

\item \textit {b jet pair mass shape for $\Z+\text{jets}$ background} \\
The uncertainty from this source is evaluated in the off-diagonal sidebands in the b jet pair mass and Z+jet mass plane, as mentioned in Section~\ref{sec:backgrounds}, and takes into account the uncertainty in the CMS cross section measurement of $\Z+\text{heavy-flavor jets}$ processes.
The resultant uncertainty in the $\Z+\text{jets}$ event yield is in the range 20 to 50\%

\item \textit {Normalization of $\Z+\text{jets}$ and $\ttbar+\text{jets}$ backgrounds} \\
The uncertainty in the event yield is $12\%$ for $\Z+\text{jets}$ and $13\%$ for $\ttbar+\text{jets}$ processes as mentioned in Section~\ref{sec:backgrounds}.

\item \textit {Diboson cross section}  \\
The uncertainty is taken from the underlying CMS diboson cross section measurements~\cite{2013PhLB..721..190C, Chatrchyan:2014aqa},
and implies a 10\% uncertainty in the estimated number of diboson events.

\end{itemize}

Sources affecting only the signal:
\begin{itemize}
\item \textit {Initial state radiation (ISR) modeling} \\
For the signal samples, the uncertainty from this source is estimated from the \pt distribution of the $\Theta^{0}\Theta^{0}$ system.
The distribution generated using \MADGRAPH is found to be in agreement with the data for heavy object systems with \pt $<$ 120\GeV \cite{sus-13-011}.
Above 120\GeV, an uncertainty is assigned that varies from  5 to 20\%, depending on the \pt.
This leads to a 1--5\% uncertainty in the yield of signal events for the mass ranges that are considered.

\item \textit {Integrated luminosity} \\
The CMS experiment collected data equivalent to 19.7\fbinv with a 2.6\% uncertainty \cite{CMS-PAS-LUM-13-001}.
This uncertainty applies only to yield of signal events.
The background events are either normalized in a control region or the normalization is taken from data.

\end{itemize}
The systematic uncertainties in event yield due to the lepton energy scale, and the lepton and jet energy resolution, are studied and found to be negligible (${<}0.1\%$) compared to the other sources of uncertainty.

\begin{table}[htp]
\centering
\topcaption{Impact of systematic uncertainties on individual event yields.  Ranges show the variation over the search regions that are considered.  Dashes indicate cases where a systematic uncertainty is not applied. Sources appearing in more than one process are treated as correlated in the limit setting.}
\label{ta:sys290}
\begin{tabular}{ l  c  c  c  c  c }
  \hline
  Source &  Signal [\%] & $\Z+\text{jets}$ [\%] & \ttbar and tW [\%] & Diboson [\%] \\
  \hline
 \multicolumn{5}{c}{ Signal and background} \\
  \hline
 b tagging  & 13--25  & 15--16  & 13--15  & 16--17 \\
 Jet energy scale & 0.2--2.6  & 6--8  & 4--6  & 5--7 \\
 Lepton ID, isolation, trigger & 0.9--1.2 & 0.9--1.2  & 0.9--1.3  & 0.9--1.2 \\
 Pileup modeling & 0.1--1.5  & 0.3--0.7  & 0.2--1.0  & 0.3--0.7 \\
  \hline
 \multicolumn{5}{c}{ Background only } \\
  \hline
 $\Z+\text{jet}$ mass shape & --- & 2--55  & ---  & --- \\
 b jet pair mass shape  & --- & 20--50  & ---  & --- \\
 Normalization & ---  & 12 & 13  & --- \\
 Diboson cross section  & --- & --- & --- & 7--10 \\
  \hline
 \multicolumn{5}{c}{ Signal only} \\
  \hline
 ISR & 1--5  & --- & --- & --- \\
 Integrated luminosity & 2.6 & --- & --- & --- \\
\hline

\end{tabular}
\end{table}

For the signal samples, the impact on the event yields due to PDF uncertainties
are estimated by following the PDF4LHC recommendation~\cite{Alekhin:2011sk,Botje:2011sn}.
These uncertainties are not used in the limit setting, but are
instead included as bands on the theoretical predictions.
Since the $\Z+\text{jets}$ and $\ttbar+\text{jets}$ backgrounds are normalized in control regions, PDF uncertainties are not applied.
Similarly the diboson+jets background normalization is taken from the data~\cite{2013PhLB..721..190C, Chatrchyan:2014aqa}.
The uncertainty in the signal yield is 7--45\%.

\section{Results}
\label{sec:result}

The numbers of observed data events in each signal region after all of the selection requirements are applied are consistent with the predictions from SM processes within two standard deviations, and are summarized in Tables~\ref{ta:result} and \ref{ta:result2}.
Table~\ref{ta:result} shows the number of events for signal, total background, and observed data, while Table~\ref{ta:result2} shows the detailed breakdown of backgrounds from different sources.

\begin{figure}[htb]
  \centering
  \includegraphics[width=0.48\textwidth]{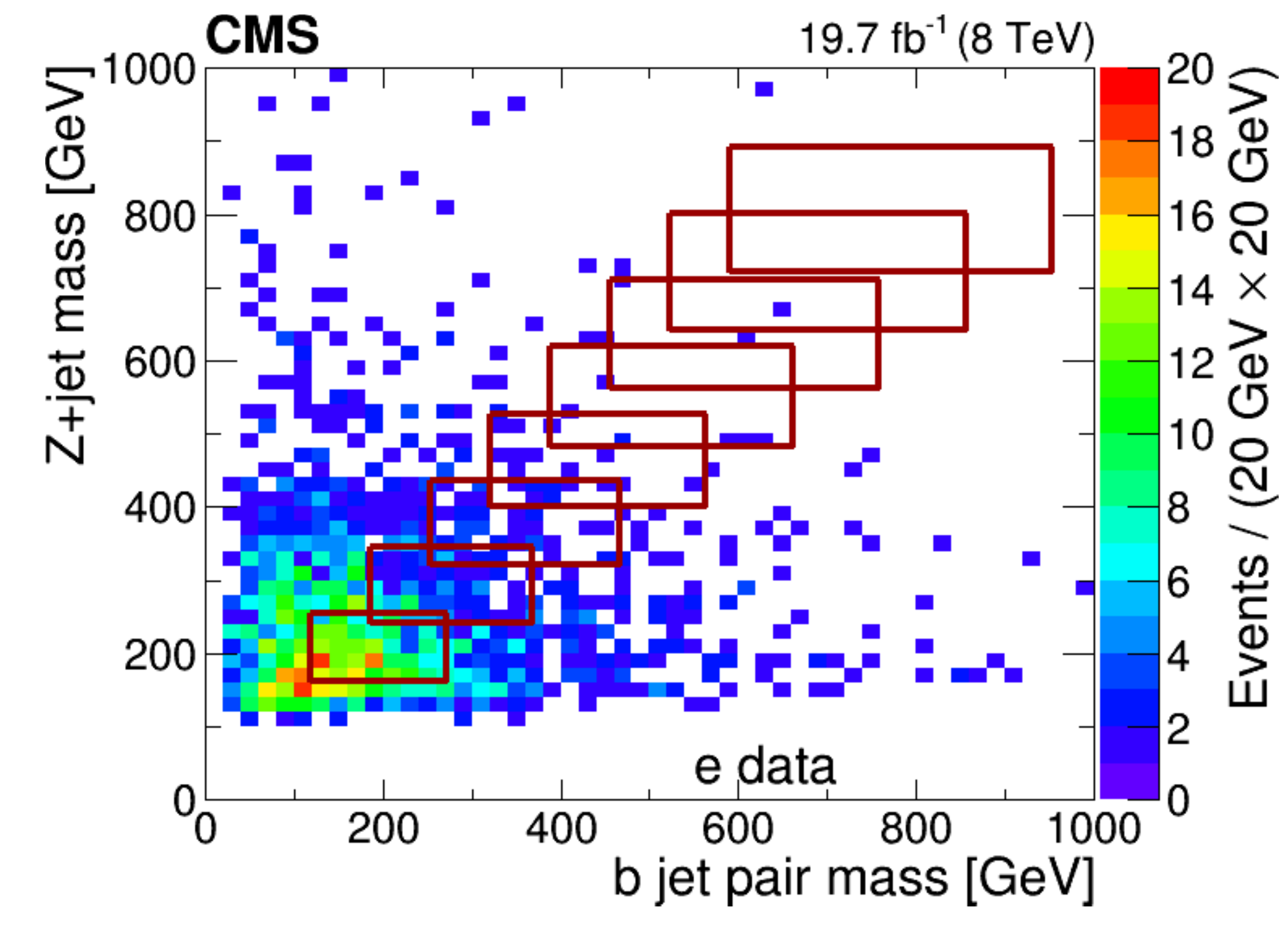}
  \includegraphics[width=0.48\textwidth]{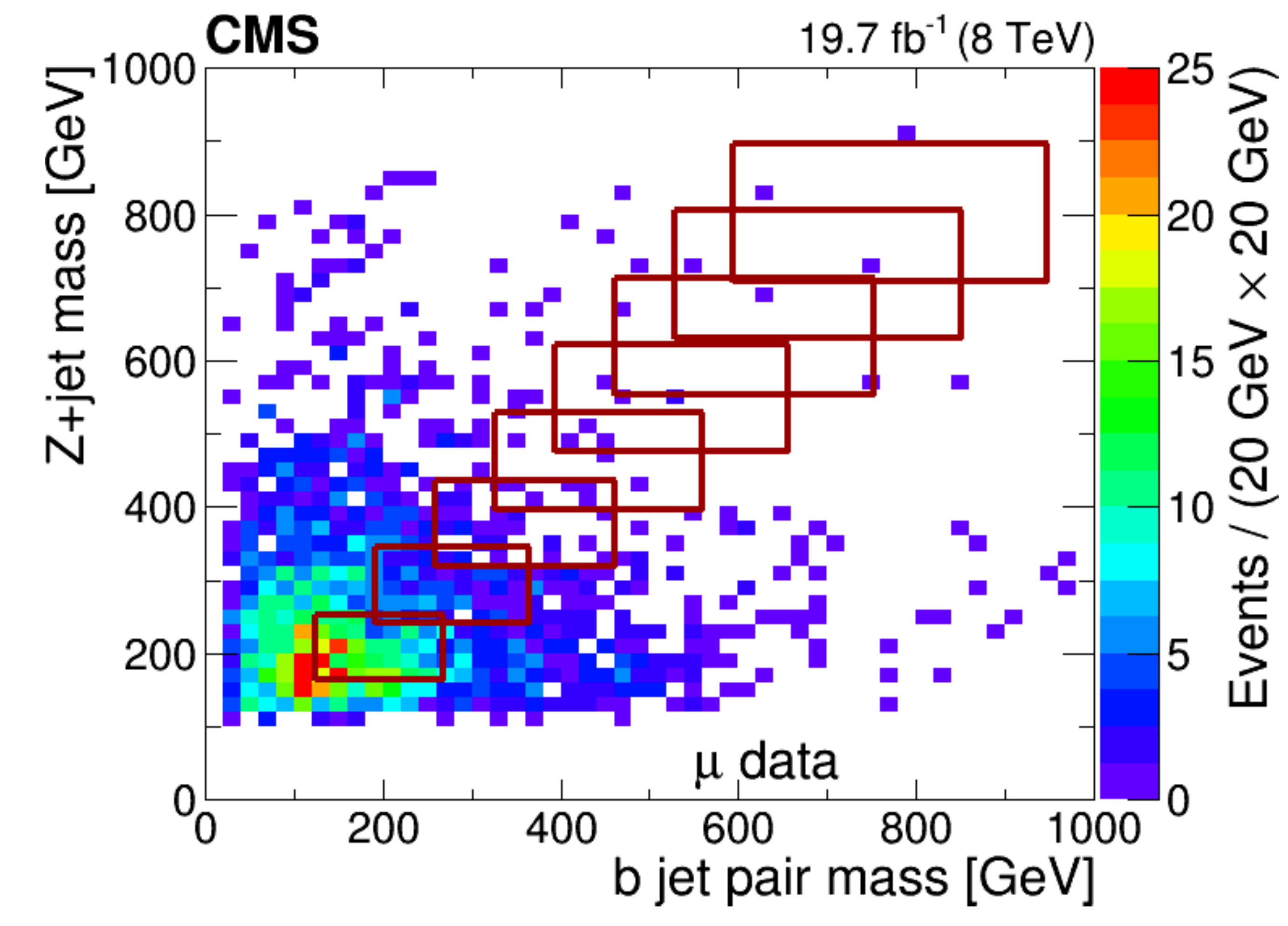}
  \includegraphics[width=0.48\textwidth]{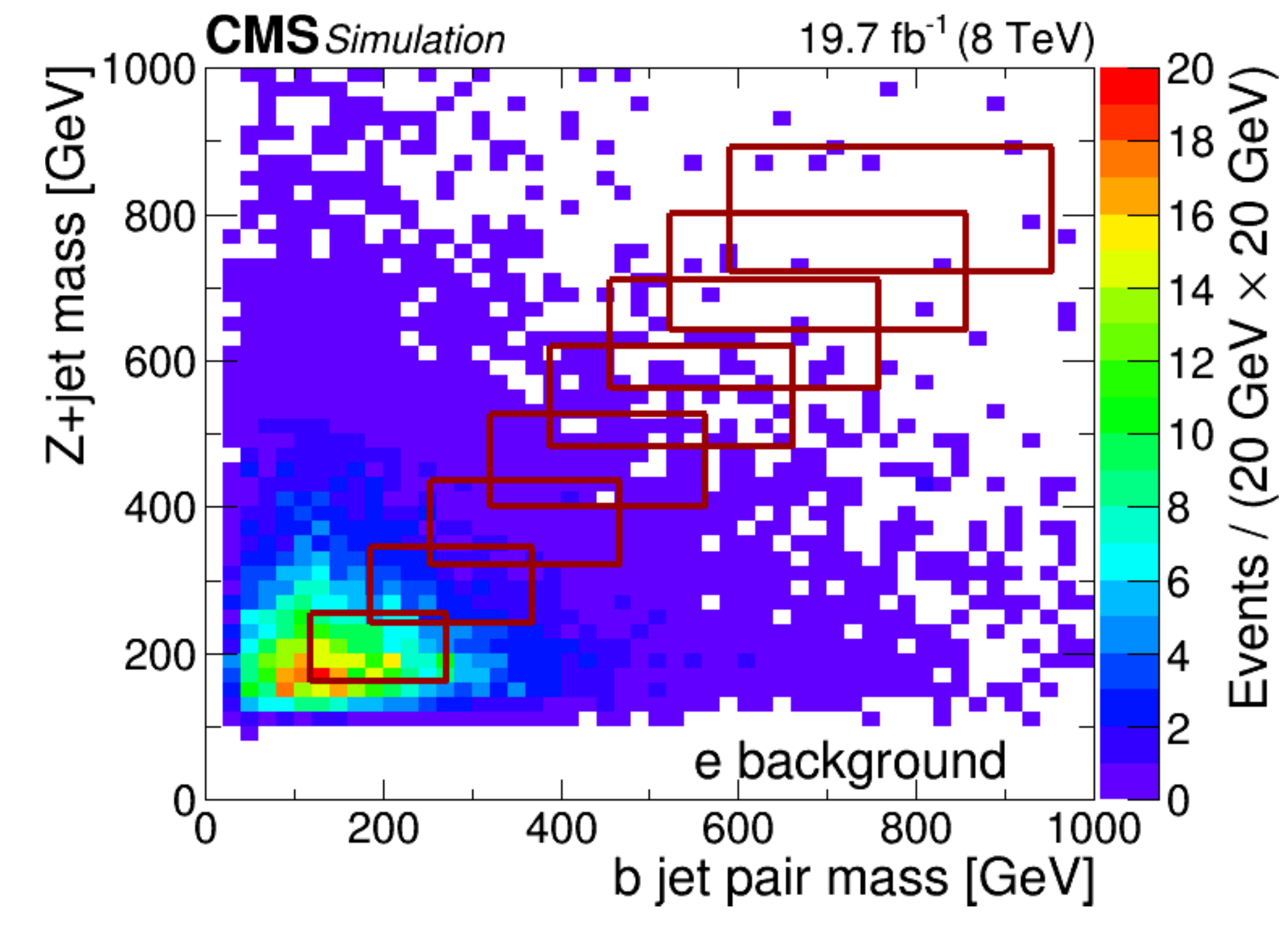}
  \includegraphics[width=0.48\textwidth]{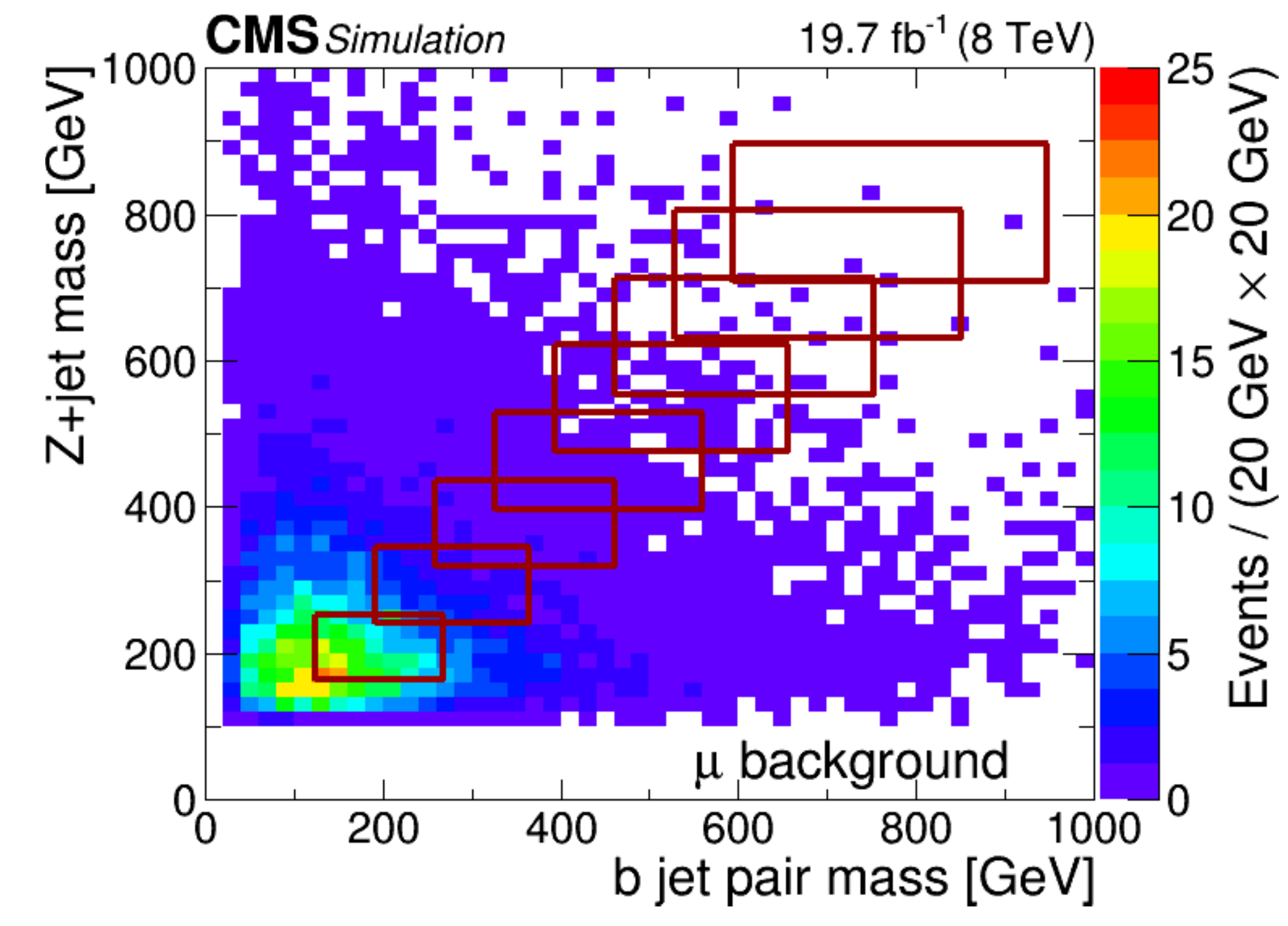}
  \caption{Distributions of the $\Z+\text{jet}$ mass versus b jet pair mass in data (top left and right), and estimated background (bottom left and right).
The number of signal candidate events is counted in the rectangular boxes defined for each signal mass hypothesis, as discussed in Section~\ref{sec:selection}.}
  \label{fig:data2D}
\end{figure}

\begin{figure}[htb]
  \centering
  \includegraphics[width=0.48\textwidth]{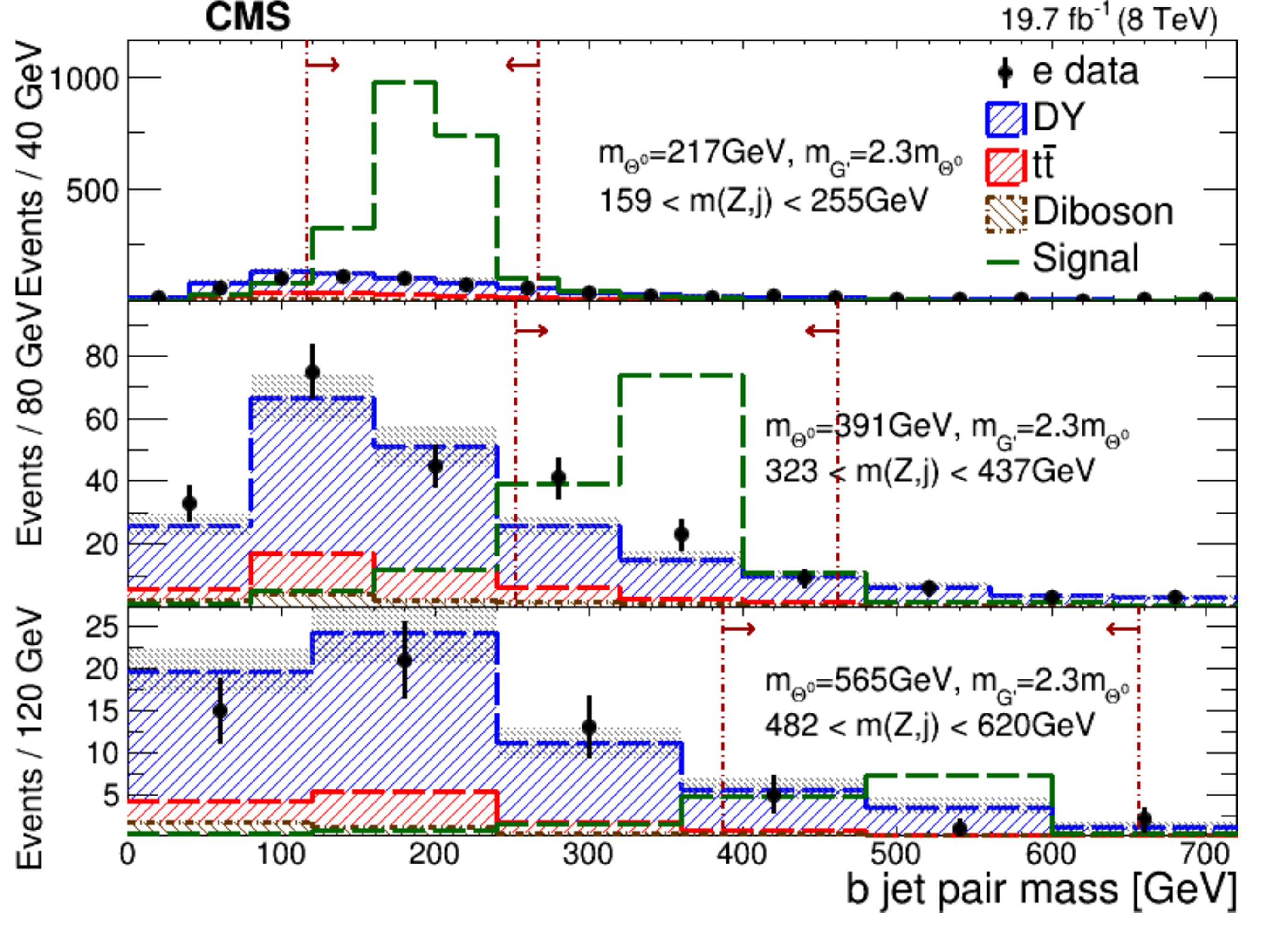}
  \includegraphics[width=0.48\textwidth]{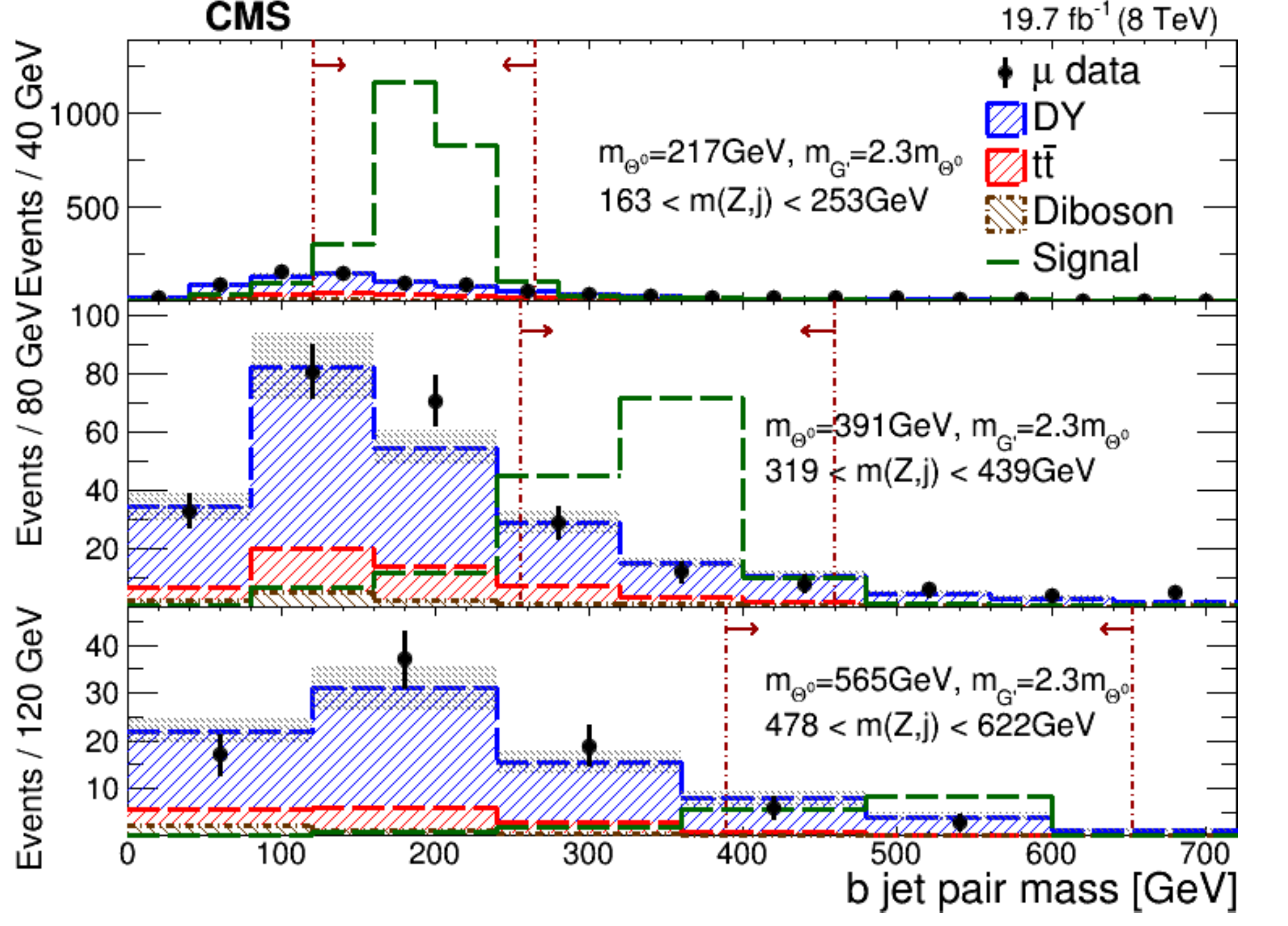}
  \includegraphics[width=0.48\textwidth]{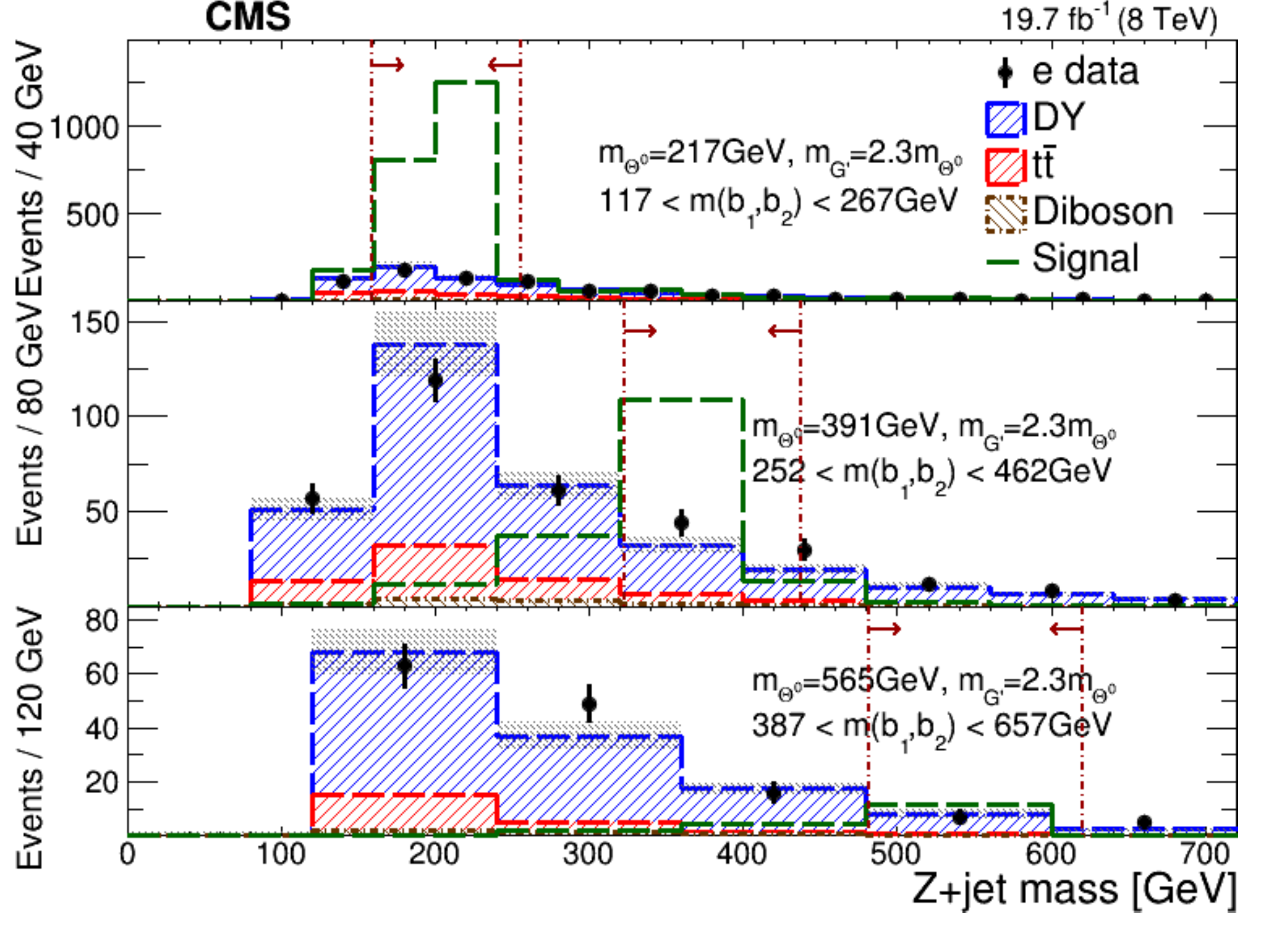}
  \includegraphics[width=0.48\textwidth]{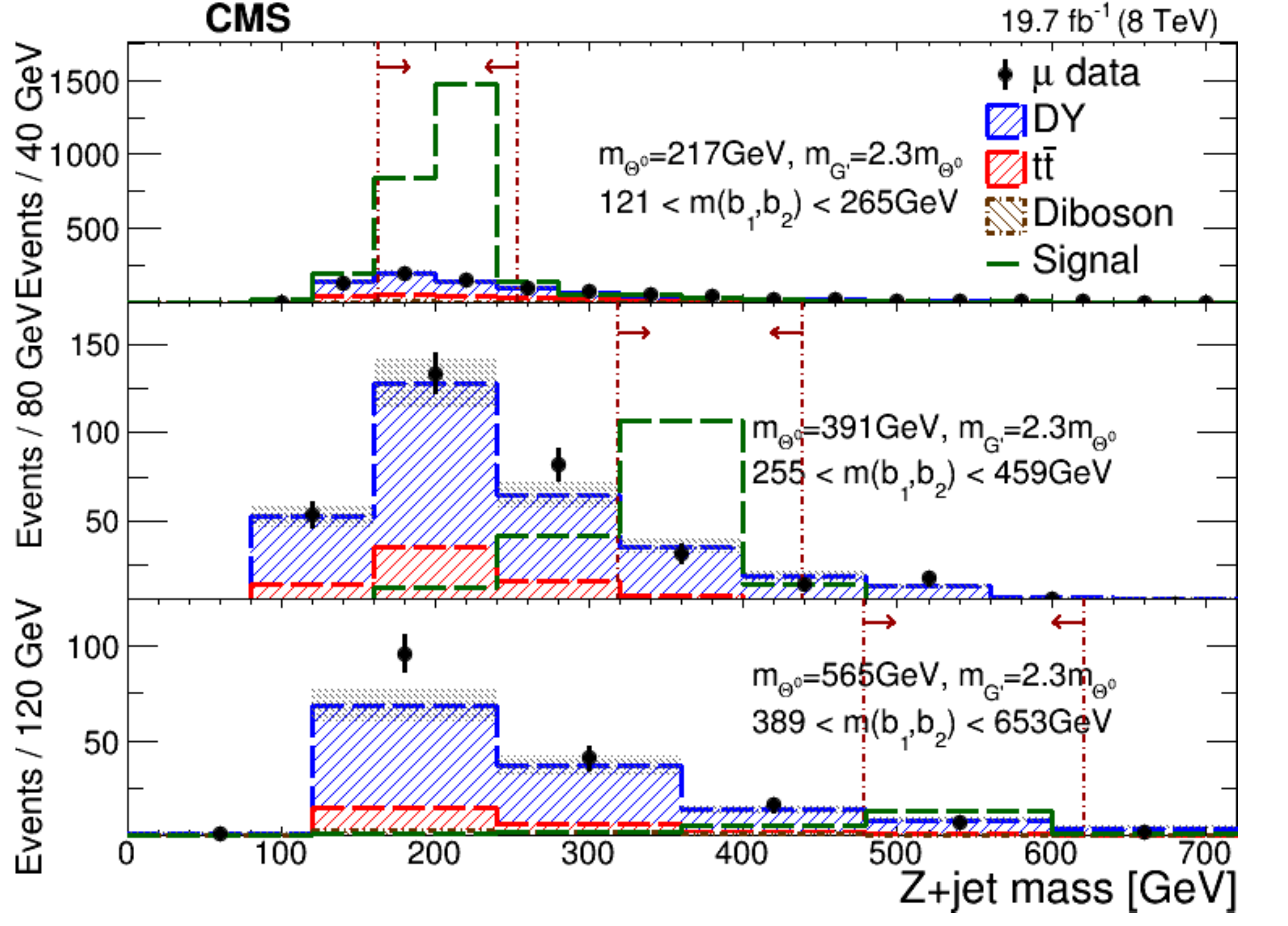}
  \caption{The b jet pair mass distributions for three different ranges of $\Z+\text{jet}$ mass in the electron channel (top left) and the muon channel (top right); $\Z+\text{jet}$ mass distributions for three different ranges of b jet pair mass in the electron channel (bottom left) and the muon channel (bottom right). The plotted regions correspond to three of the search regions.
Predicted signal distributions are overlaid.
The shaded band represents the statistical uncertainty combined with the systematic uncertainty in the simulated samples.}
  \label{fig:dataSlice}
\end{figure}

Events observed in the b jet pair and $\Z+\text{jet}$ mass plane, together with the predictions from
background processes, are plotted in Fig.~\ref{fig:data2D}.
The b jet pair and $\Z+\text{jet}$ mass distributions are shown separately in Fig.~\ref{fig:dataSlice}.

\begin{figure}[htb]
  \centering
  \includegraphics[width=0.48\textwidth]{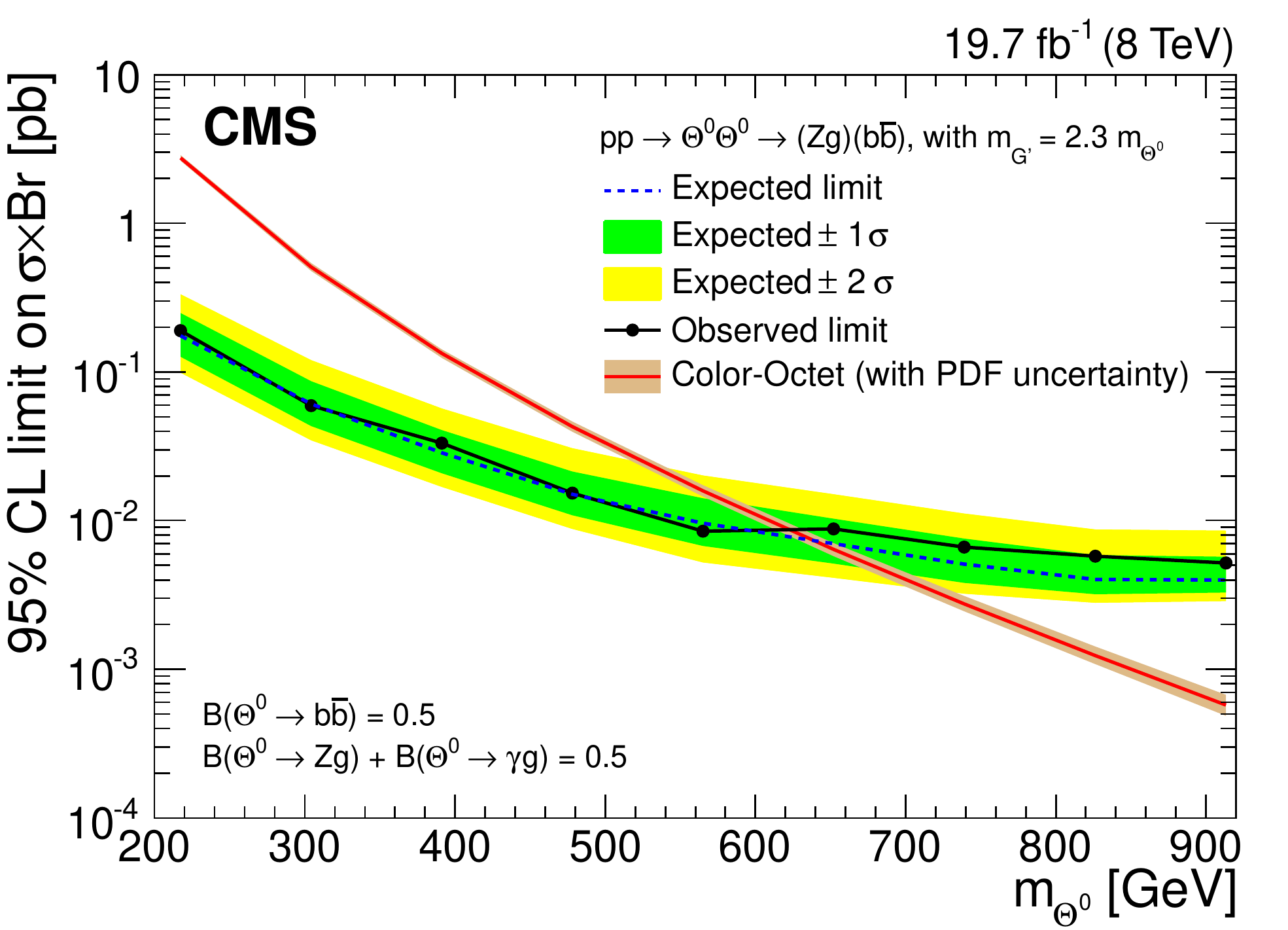}
  \includegraphics[width=0.48\textwidth]{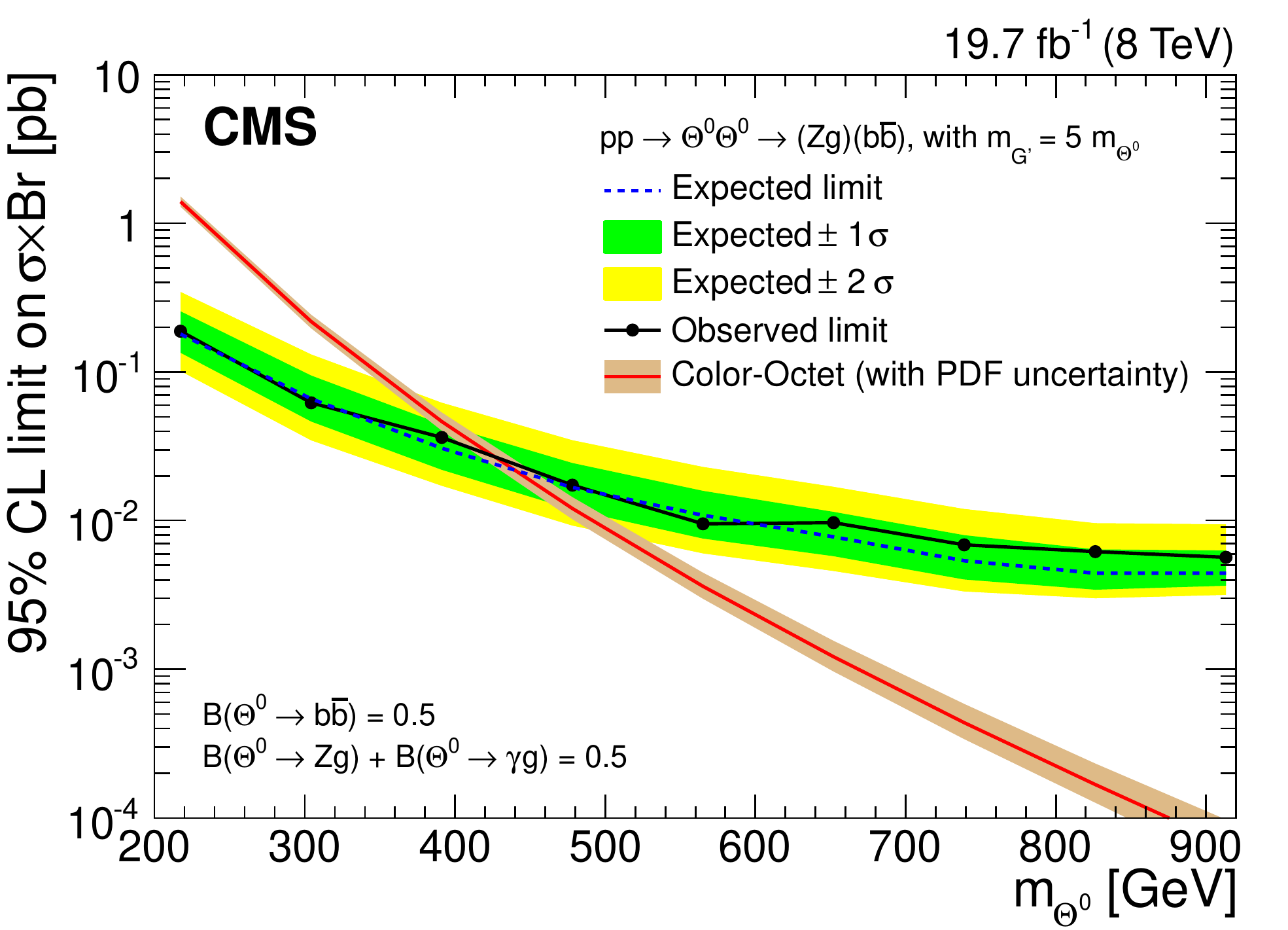}
  \caption{The 95\% CL expected and observed upper limits on the cross section times branching fraction,
$\sigma \times \mathcal{B}(\Theta^{0} \to Z \Glu ) \times \mathcal{B}(Z \to \ell\ell) \times  \mathcal{B}(\Theta^{0} \to \bbbar) \times 2$,
as a function of $\Theta^{0}$ mass, for the case where $m_{\mathrm{G'}} = 2.3 m_{\Theta^{0}}$ (left) and $m_{\mathrm{G'}} = 5 m_{\Theta^{0}}$ (right) with the band on the color octet theoretical prediction
indicating uncertainty in the signal yield due to PDF.
These results assume  $\mathcal{B}$($\Theta^{0} \to \bbbar$) = 0.5 and $\mathcal{B}$($\Theta^{0} \to \Z\Glu$) + $\mathcal{B}$($\Theta^{0} \to \gamma\Glu$) = 0.5.
}
  \label{fig:limit_r23_and_r5}
\end{figure}

Upper limits are calculated on the ratio of the measured $\Theta^{0}$ pair production cross section times branching fraction
$\mathcal{B}(\Theta^{0} \to Z \Glu ) \times \mathcal{B}(Z \to \ell\ell) \times \mathcal{B}(\Theta^{0} \to \bbbar) \times 2$, to the theoretically expected cross section times branching fraction, for each search region separately.
We have used as a benchmark the case where $\mathcal{B}(\Theta^{0} \to \bbbar) = \mathcal{B}(\Theta^{0} \to Z\Glu) +
\mathcal{B}(\Theta^{0} \to \gamma\Glu)= 0.5$.
The resulting limits placed on octo-triplet particle production cross section and mass can be scaled
for different choices of $\mathcal{B}(\Theta^{0} \to \bbbar)$ and $\mathcal{B}(\Theta^{0} \to \Z \Glu)$.
The results in the electron and muon channels are consistent and therefore combined.
A modified-frequentist approach, $CL_{s}$, is used to calculate the limits
\cite{cls-technique, 1999NIMPA.434..435J}.
For each mass point, pseudo-experiments are run for the signal plus background hypothesis, and for the background only hypothesis.
The systematic uncertainties discussed in Section~\ref{sec:systematics}, and their correlations,
are included in the limit determinations through a set of nuisance parameters.
The results are shown, in Fig.~\ref{fig:limit_r23_and_r5}, as a 95\% confidence level (CL) observed and expected limit on
$\sigma \times \mathcal{B}(\Theta^{0} \to Z \Glu ) \times \mathcal{B}(Z \to \ell\ell) \times  \mathcal{B}(\Theta^{0} \to \bbbar) \times 2$, as a function of $\Theta^{0}$ mass.
For the case where $m_{\mathrm{G'}} = 2.3 m_{\Theta^{0}}$, this result excludes $\Theta^{0}$ masses
below $623\GeV$ at 95\% CL, with an expected exclusion of $639\GeV$.
For the case where $m_{\mathrm{G'}} = 5 m_{\Theta^{0}}$, $\Theta^{0}$ masses
below $426\GeV$ are excluded at 95\% CL, with an expected exclusion of 439\GeV.
These plots also include the theoretical predictions, with the band indicating the uncertainty in the signal due to the PDF uncertainty.

\begin{table}[htb]
\centering
\topcaption{The number of events after final selection for the signal, total background, and observed data, together with the statistical and systematic uncertainty. For the entries where a single uncertainty is shown, the statistical uncertainty is negligible.}
\label{ta:result}
{
\begin{tabular}{  c | c  c  c  c }
  \hline
 \multicolumn{1}{c}{$m_{\Theta^{0}}$ [\GeVns{}]} &  \multicolumn{2}{c}{Signal} & Background & Observed \\ \cline{2-3}
\multicolumn{1}{c}{} & $m_{\mathrm{G'}} = 2.3 m_{\Theta^{0}}$ & $m_{\mathrm{G'}} = 5 m_{\Theta^{0}}$ & & \\
 \hline
 \multicolumn{5}{c}{Electron channel} \\
 \hline
217 &  2110${\pm}$70${\pm}$290  &  1110${\pm}$40${\pm}$160  &   358${\pm}$15${\pm}$81  &  336  \\
304 &  448${\pm}$12${\pm}$70  &  188${\pm}$6${\pm}$31  &   123${\pm}$6${\pm}$27  &  115  \\
391 &  119${\pm}$3${\pm}$22  &  38.6${\pm}$1.3${\pm}$7.4  &   42.3${\pm}$3.0${\pm}$11.0  &  59  \\
478 &  35.8${\pm}$1.0${\pm}$6.6  &  9.45${\pm}$0.32${\pm}$1.90  &   18.9${\pm}$2.7${\pm}$6.4  &  24  \\
565 &  11.7${\pm}$0.3${\pm}$2.4  &  2.37${\pm}$0.09${\pm}$0.53  &   8.14${\pm}$1.94${\pm}$3.40  &  7  \\
652 &  3.96${\pm}$0.13${\pm}$0.86  &  0.72${\pm}$0.03${\pm}$0.17  &   3.35${\pm}$0.82${\pm}$1.50  &  4  \\
739 &  1.42${\pm}$0.06${\pm}$0.32  &  0.23${\pm}$0.01${\pm}$0.05  &   1.02${\pm}$0.31${\pm}$0.51  &  1  \\
826 &  0.56${\pm}$0.02${\pm}$0.13  &  0.08${\pm}$             0.02  &   0.66${\pm}$0.33${\pm}$0.36  &  0  \\
913 &  0.24${\pm}$0.01${\pm}$0.06  &  0.03${\pm}$             0.01  &   0.31${\pm}$0.14${\pm}$0.19  &  0  \\
\hline
 \multicolumn{5}{c}{Muon channel} \\
 \hline
217 &  2360${\pm}$70${\pm}$310  &  1170${\pm}$40${\pm}$160  &   348${\pm}$11${\pm}$74  &  355  \\
304 &  486${\pm}$12${\pm}$74  &  185${\pm}$6${\pm}$31  &   126${\pm}$7${\pm}$27 &  127  \\
391 &  118${\pm}$3${\pm}$20  &  38.7${\pm}$1.2${\pm}$7.3  &   44.8${\pm}$3.1${\pm}$11.3  &  39  \\
478 &  38.7${\pm}$1.0${\pm}$7.4  &  9.58${\pm}$0.32${\pm}$1.90  &   17.2${\pm}$1.9${\pm}$5.7  &  15  \\
565 &  13.1${\pm}$0.4${\pm}$2.7  &  2.70${\pm}$0.09${\pm}$0.58  &   9.52${\pm}$1.40${\pm}$3.80  &  8  \\
652 &  4.58${\pm}$0.14${\pm}$0.99  &  0.74${\pm}$0.03${\pm}$0.17  &   3.25${\pm}$0.84${\pm}$1.30  &  5  \\
739 &  1.63${\pm}$0.07${\pm}$0.36  &  0.24${\pm}$0.01${\pm}$0.06  &   1.48${\pm}$0.49${\pm}$0.58  &  3  \\
826 &  0.69${\pm}$0.03${\pm}$0.16  &  0.08${\pm}$             0.02  &   0.32${\pm}$0.17${\pm}$0.14  &  2  \\
913 &  0.26${\pm}$0.01${\pm}$0.06  &  0.03${\pm}$             0.01  &   0.14${\pm}$0.14${\pm}$0.09  &  1  \\
  \hline
\end{tabular}
}
\end{table}

\begin{table}[htp]
\centering

\topcaption{The number of background events, from all sources, after final selection, together with the statistical and systematic uncertainty. For the entries where a single uncertainty is shown, the systematic uncertainty is negligible.}
\label{ta:result2}
{
\begin{tabular}{  c | c  c  c }
  \hline
 $m_{\Theta^{0}}$ [\GeVns{}] & $\Z+\text{jets}$ & \ttbar and tW & Diboson  \\
  \hline

 \multicolumn{4}{c}{Electron channel} \\

 \hline
217 &   256${\pm}$15${\pm}$79  &   91.3${\pm}$2.2${\pm}$18.0  &   11.1${\pm}$0.9${\pm}$2.2  \\
304 &   90.5${\pm}$5.7${\pm}$26.0  &   27.5${\pm}$1.2${\pm}$5.3  &   4.55${\pm}$0.60${\pm}$0.90  \\
391 &   33.6${\pm}$2.9${\pm}$11.0  &   6.55${\pm}$0.60${\pm}$1.30  &   2.10${\pm}$0.35${\pm}$0.42  \\
478 &   16.9${\pm}$2.7${\pm}$6.4  &   1.24${\pm}$0.25${\pm}$0.24  &   0.78${\pm}$0.21${\pm}$0.16  \\
565 &   7.36${\pm}$1.90${\pm}$3.30  &   0.36${\pm}$0.11${\pm}$0.07  &   0.42${\pm}$0.17${\pm}$0.08  \\
652 &   3.08${\pm}$0.81${\pm}$1.50  &   0.04${\pm}$0.03${\pm}$0.01  &   0.23${\pm}$0.12${\pm}$0.05  \\
739 &   1.02${\pm}$0.31${\pm}$0.51  &   ${<}0.04$                          &   ${<}0.01$   \\
826 &   0.65${\pm}$0.33${\pm}$0.36  &   ${<}0.04$                          &   0.01${\pm}$0.01  \\
913 &   0.27${\pm}$0.13${\pm}$0.19  &   ${<}0.04$                          &   0.04${\pm}$0.04${\pm}$0.01  \\
\hline
 \multicolumn{4}{c}{Muon channel} \\
 \hline
217 &   244${\pm}$10${\pm}$72  &   91.5${\pm}$2.2${\pm}$18.0  &   12.7${\pm}$1.0${\pm}$2.4  \\
304 &   92.2${\pm}$7.2${\pm}$27.0 &   28.7${\pm}$1.2${\pm}$5.5  &   5.57${\pm}$0.59${\pm}$1.10  \\
391 &   35.1${\pm}$3.0${\pm}$11.0  &   7.51${\pm}$0.62${\pm}$1.40  &   2.20${\pm}$0.36${\pm}$0.42  \\
478 &   15.0${\pm}$1.8${\pm}$5.7  &   1.20${\pm}$0.26${\pm}$0.23  &   1.00${\pm}$0.24${\pm}$0.19  \\
565 &   8.89${\pm}$1.40${\pm}$3.80  &   0.36${\pm}$0.13${\pm}$0.07  &   0.26${\pm}$0.11${\pm}$0.05  \\
652 &   3.03${\pm}$0.84${\pm}$1.30  &   0.08${\pm}$0.07${\pm}$0.01  &   0.14${\pm}$0.05${\pm}$0.03  \\
739 &   1.27${\pm}$0.48${\pm}$0.58  &   0.12${\pm}$0.07${\pm}$0.02  &   0.09${\pm}$0.04${\pm}$0.02  \\
826 &   0.23${\pm}$0.16${\pm}$0.14  &   0.05${\pm}$0.05${\pm}$0.01  &   0.05${\pm}$0.03${\pm}$0.01  \\
913 &   0.14${\pm}$0.14${\pm}$0.09  &   ${<}0.05$                           &   ${<}0.05$  \\
  \hline
\end{tabular}
}
\end{table}

\clearpage
\section{Summary}\label{sec:conclusion}
A search for pair production of neutral color-octet weak-triplet scalar particles ($\Theta^{0}$) has been performed
based on processes where one $\Theta^{0}$ decays to a pair of b quark jets and the other to a \Z boson plus a jet, with
the \Z boson decaying to a pair of electrons or muons.
This analysis is based on data collected with the CMS experiment in proton-proton collisions at $\sqrt{s}$ = 8\TeV, corresponding to an integrated luminosity of 19.7\fbinv.
The number of observed events is found to be in agreement with the standard model prediction.
The $CL_{s}$ method is used to set a 95\% confidence level limit on the cross section of
octo-triplet particles, assuming $\mathcal{B}(\Theta^{0} \to \bbbar) = 0.5,$ with the remaining $\Theta^{0}$ branching fraction shared between $\Z\Glu$ and $\gamma\Glu$.
By comparing the theoretical predictions of the octo-triplet model and the observed limits,
masses of $\Theta^{0}$ below $623\GeV$ for $m_{\mathrm{G'}} = 2.3 m_{\Theta^{0}}$,
and below $426\GeV$ for $m_{\mathrm{G'}} = 5 m_{\Theta^{0}}$, are excluded at 95\% confidence level.
These are the first direct experimental bounds on the $\Theta^{0}$ production model.

\begin{acknowledgments}\label{sec:Acknowledgments}
We congratulate our colleagues in the CERN accelerator departments for the excellent performance of the LHC and thank the technical and administrative staffs at CERN and at other CMS institutes for their contributions to the success of the CMS effort. In addition, we gratefully acknowledge the computing centers and personnel of the Worldwide LHC Computing Grid for delivering so effectively the computing infrastructure essential to our analyses. Finally, we acknowledge the enduring support for the construction and operation of the LHC and the CMS detector provided by the following funding agencies: BMWFW and FWF (Austria); FNRS and FWO (Belgium); CNPq, CAPES, FAPERJ, and FAPESP (Brazil); MES (Bulgaria); CERN; CAS, MoST, and NSFC (China); COLCIENCIAS (Colombia); MSES and CSF (Croatia); RPF (Cyprus); MoER, ERC IUT and ERDF (Estonia); Academy of Finland, MEC, and HIP (Finland); CEA and CNRS/IN2P3 (France); BMBF, DFG, and HGF (Germany); GSRT (Greece); OTKA and NIH (Hungary); DAE and DST (India); IPM (Iran); SFI (Ireland); INFN (Italy); MSIP and NRF (Republic of Korea); LAS (Lithuania); MOE and UM (Malaysia); CINVESTAV, CONACYT, SEP, and UASLP-FAI (Mexico); MBIE (New Zealand); PAEC (Pakistan); MSHE and NSC (Poland); FCT (Portugal); JINR (Dubna); MON, RosAtom, RAS and RFBR (Russia); MESTD (Serbia); SEIDI and CPAN (Spain); Swiss Funding Agencies (Switzerland); MST (Taipei); ThEPCenter, IPST, STAR and NSTDA (Thailand); TUBITAK and TAEK (Turkey); NASU and SFFR (Ukraine); STFC (United Kingdom); DOE and NSF (USA).

Individuals have received support from the Marie-Curie program and the European Research Council and EPLANET (European Union); the Leventis Foundation; the A. P. Sloan Foundation; the Alexander von Humboldt Foundation; the Belgian Federal Science Policy Office; the Fonds pour la Formation \`a la Recherche dans l'Industrie et dans l'Agriculture (FRIA-Belgium); the Agentschap voor Innovatie door Wetenschap en Technologie (IWT-Belgium); the Ministry of Education, Youth and Sports (MEYS) of the Czech Republic; the Council of Science and Industrial Research, India; the HOMING PLUS program of the Foundation for Polish Science, cofinanced from European Union, Regional Development Fund; the Compagnia di San Paolo (Torino); the Consorzio per la Fisica (Trieste); MIUR project 20108T4XTM (Italy); the Thalis and Aristeia programs cofinanced by EU-ESF and the Greek NSRF; and the National Priorities Research Program by Qatar National Research Fund.
\end{acknowledgments}
\newpage

\bibliography{auto_generated}
\cleardoublepage \appendix\section{The CMS Collaboration \label{app:collab}}\begin{sloppypar}\hyphenpenalty=5000\widowpenalty=500\clubpenalty=5000\textbf{Yerevan Physics Institute,  Yerevan,  Armenia}\\*[0pt]
V.~Khachatryan, A.M.~Sirunyan, A.~Tumasyan
\vskip\cmsinstskip
\textbf{Institut f\"{u}r Hochenergiephysik der OeAW,  Wien,  Austria}\\*[0pt]
W.~Adam, E.~Asilar, T.~Bergauer, J.~Brandstetter, E.~Brondolin, M.~Dragicevic, J.~Er\"{o}, M.~Flechl, M.~Friedl, R.~Fr\"{u}hwirth\cmsAuthorMark{1}, V.M.~Ghete, C.~Hartl, N.~H\"{o}rmann, J.~Hrubec, M.~Jeitler\cmsAuthorMark{1}, V.~Kn\"{u}nz, A.~K\"{o}nig, M.~Krammer\cmsAuthorMark{1}, I.~Kr\"{a}tschmer, D.~Liko, T.~Matsushita, I.~Mikulec, D.~Rabady\cmsAuthorMark{2}, B.~Rahbaran, H.~Rohringer, J.~Schieck\cmsAuthorMark{1}, R.~Sch\"{o}fbeck, J.~Strauss, W.~Treberer-Treberspurg, W.~Waltenberger, C.-E.~Wulz\cmsAuthorMark{1}
% ...
\vskip\cmsinstskip
% ...
1:~~Also at Vienna University of Technology, Vienna, Austria\\
2:~~Also at CERN, European Organization for Nuclear Research, Geneva, Switzerland\\
% ...

\end{sloppypar}
\end{document}